\documentclass[%
 reprint,
superscriptaddress,
 amsmath,amssymb,
 aps,
pra,
]{revtex4-2}
\usepackage{graphicx}
\usepackage{dcolumn}
\usepackage{bm}
\usepackage{nicefrac}
\usepackage{bbold}
\usepackage{amsmath}
\usepackage{braket}
\usepackage{siunitx}
\usepackage[dvipsnames]{xcolor}
\usepackage{appendix}
\begin{document}
%
%
\title{Optics-microwave entanglement and state teleportation mediated by a cavity magnomechanical system}
\author{F. Engelhardt}
\email{engelhardt@physik.rwth-aachen.de}
\affiliation{Institute for Theoretical Solid State Physics, RWTH Aachen University, 52074
Aachen, Germany}
\affiliation{Max Planck Institute for the Science of Light, Staudtstr. 2, PLZ 91058 Erlangen, Germany}
\author{A. V. Bondarenko}
\affiliation{Kavli Institute of Nanoscience, Delft University of Technology, Delft, 2628CJ, the Netherlands}
\affiliation{Institute of Magnetism of NAS and MOS of Ukraine, Kyiv, 03142, Ukraine}
\author{A. Metelmann}
\affiliation{Institute for Theory of Condensed Matter and Institute for Quantum Materials and Technology, Karlsruhe Institute of Technology, 76131, Karlsruhe, Germany}
\affiliation{ISIS (UMR 7006), Université de Strasbourg, 67000 Strasbourg, France}
\author{Ya. M. Blanter}
\affiliation{Kavli Institute of Nanoscience, Delft University of Technology, Delft, 2628CJ, the Netherlands}
\author{S. {Viola Kusminskiy}}
\affiliation{Institute for Theoretical Solid State Physics, RWTH Aachen University, 52074
Aachen, Germany}
\affiliation{Max Planck Institute for the Science of Light, Staudtstr. 2, PLZ 91058 Erlangen, Germany}
\author{V. A. S. V. Bittencourt}
\email{sant@unistra.fr}
\affiliation{ISIS (UMR 7006), Université de Strasbourg, 67000 Strasbourg, France}
\date{\today}
\begin{abstract}
Generating usable output-entanglement in continuous variable systems can serve as a viable resource for improving applications in quantum information science. In this work, we show how to generate steady-state output-entanglement in a  two-stage conversion setup between optical and microwave photon which employs resonantly coupled magnetic and mechanical excitations, as proposed in Phys. Rev. Applied 18, 044059 (2022). We show that the entanglement can be maximized for the same set of parameters which optimize the frequency-conversion efficiency, and that it can be leveraged for a teleportation-based state-transfer protocol for coherent input-states with fidelity close to unity. We propose an implementation based on an Yittrium Iron Garnet disk of micrometer scale, and use both simulation results and reasonable estimates to assess the performance under optimized conditions. We find a maximum teleportation fidelity of $0.75$ for the proposed setup.
\end{abstract}
\maketitle
\section{Introduction}
Quantum entanglement is central to applications in quantum technologies, including cryptography~\cite{yinEntanglementbasedSecureQuantum2020,ekertQuantumCryptographyBased1991,bennettQuantumCryptographyBells1992, tittelQuantumCryptographyUsing2000,bennettQuantumCryptographyUsing1992}, teleportation~\cite{bennettTeleportingUnknownQuantum1993, furusawaUnconditionalQuantumTeleportation1998, boschiExperimentalRealizationTeleporting1998,vaidmanTeleportationQuantumStates1994, braunsteinTeleportationContinuousQuantum1998, bouwmeesterExperimentalQuantumTeleportation1997,riebeDeterministicQuantumTeleportation2004, barrettDeterministicQuantumTeleportation2004}, sensing, and quantum networks~\cite{vanloockMultipartiteEntanglementContinuous2000, jiangDistributedQuantumComputation2007, pirandolaPhysicsUniteBuild2016, zhuangDistributedQuantumSensing2018,guoDistributedQuantumSensing2020,zhangEntanglementEnhancedSensingLossy2015,xiaEntanglementenhancedOptomechanicalSensing2023,ciracQuantumStateTransfer1997, acinEntanglementPercolationQuantum2007,caleffiQuantumTransductionEnabling2026}.
Generating entanglement is, nevertheless, not a trivial task, in particular when the entangled degrees-of-freedom belong to systems operating at different frequencies, as it is the case for optical and microwave photons. While microwave photons can couple resonantly to superconducting qubits, a lead architecture for quantum computation~\cite{aruteQuantumSupremacyUsing2019}, optical photons operate at much higher frequency and are ideal for transporting quantum information over long distances at low losses~\cite{magnardMicrowaveQuantumLink2020}. Entangling optical and microwave photons can thus enhance capabilities of quantum networks for information transfer and processing~\cite{laukPerspectivesQuantumTransduction2020}.\\
\begin{figure}
    \centering
    \includegraphics[width = 0.48\textwidth]{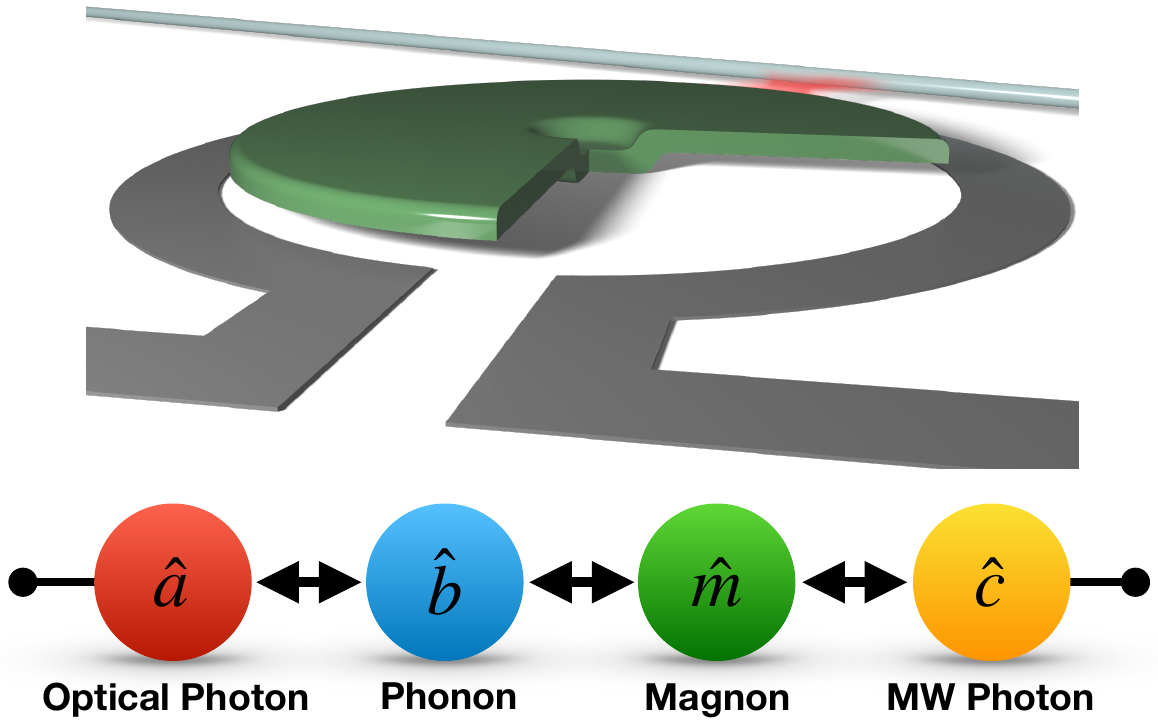}
    \caption{3D render of the proposed geometry. A telecom-frequency light field inside an optical fiber couples evanescently to a micromagnetic disk. The disk serves as optical cavity while simultaneously hosting mechanical and magnetic modes as well. The last element is a dark-grey omega-shaped microwave ring resonator. In our simulations, the disk has radius $3.7\,\rm{\mu m}$ and thickness $200\,\rm{n m}$. The envisioned coupling scheme is pictured below the setup.}
    \label{Fig:Intro}
\end{figure}
This can be achieved, for example, via direct electro-optic interaction~\cite{ruedaElectroopticEntanglementSource2019,zhangGenerationGeneralizedHybrid2021,tsangCavityQuantumElectrooptics2011,davossaQuantumInternetEntanglement2023}, which was used in a recent seminal experimental demonstration~\cite{sahuEntanglingMicrowavesLight2023}. Alternatively, one can use a mediator capable of coupling to both types of excitations, lifting some limitations. A prominent example are mechanical degrees of freedom, which can mediate engineered interactions between itinerant and output modes~\cite{andrewsBidirectionalEfficientConversion2014,miaoQuantumPositioningScheme2024,renNonreciprocalOpticalMicrowave2022,shangGenerationMicrowaveopticsEntanglement2024,zhengNonreciprocalMicrowaveopticalEntanglement2024,jiaEntanglementMicrowaveFields2025,meesalaQuantumEntanglementOptical2024,zhongEntanglementMicrowaveopticalModes2020,zhongProposalHeraldedGeneration2020,zhongMicrowaveOpticalEntanglement2022,zhongMicrowaveopticalEntanglementPulsepumped2024,weiTunableMicrowaveopticalEntanglement2022,rauEntanglementThresholdsDoubly2022}. Efficient coupling to photons can be achieved by hosting microwave-frequency phonons in an optical and microwave cavity, which enhances the coupling strength due to the confinement of photons. The radiation-pressure interaction between optical photons and phonons allows the activation of a photon-phonon pair creation process via an optical drive set at a frequency higher from the cavity by one phonon frequency (blue detuning). The same driving scheme can, in principle, also be applied to a system that uses magnetic degrees of freedom as mediator via magneto-optical interactions~\cite{liMagnonPhotonPhononEntanglementCavity2018}. In contrast to phonons, whose frequencies are generally fixed by the resonator design, the magnetic excitations (magnons) are easily tunable via an external magnetic field, which adds flexibility in design and makes them a promising new candidate for applications in quantum information processing~\cite{chumakMagnonSpintronics2015,chumakAdvancesMagneticsRoadmap2022,lachance-quirionHybridQuantumSystems2019,flebus2024MagnonicsRoadmap2024}.\\ 
The capacity of a setup based on hybrid quantum systems to generate and distribute entanglement is generally dependent on how strong the subsystems interact. Together with taking into account their individual lifetimes, this can be quantified by the cooperativity. For two systems A and B, the cooperativty is defined as $C_{AB}=4g_{AB}^2 /\gamma_{A,\rm{tot}} \gamma_{B,\rm{tot}}$, where $g_{AB}$ is the coupling strength and $\gamma_{A/B,{\rm{tot}}}$ the respective total linewidth. A setup with one mediator between two output modes is thereby quantified by two cooperativities. It has been shown that for one-stage setups~\cite{wangBipartiteTripartiteOutput2015} maximum output entanglement occurs for matching cooperativities. This condition is one of the main bottlenecks for one-stage setups, therefore devices bridging optics and microwaves would benefit a lot from added tunability, which further gives rise to more flexibility in design and functionality. This can be achieved by combining magnons and phonons and using their strong resonant interaction~\cite{anCoherentLongrangeTransfer2020,mullerChiralPhononsPhononic2024} as mediator. In essence, the magnomechanical interaction can be used as a two-stage mediator for optics-to-microwave frequency conversion, in which cooperativity constraints typically encountered on single-stage converters are lifted~\cite{engelhardtOptimalBroadbandFrequency2022}.\\

In this work, we investigate the optimal conditions for generating entanglement between an output optical mode and an output microwave mode which interact via a hybrid magnomechanical system shown in Fig.~\ref{Fig:Intro}: a magnetic dielectric that simultaneously serves as an optical cavity while also hosting the mechanical mode, and is coupled to a microwave resonator. The optical and microwave modes further couple to external ports, which can be, for example, an optical fiber and a microwave cable, respectively. The optical mode is driven with a blue-detuned coherent drive, which generates entanglement between optical photons and phonons. The interaction between phonons and magnons, and between magnons and microwaves, distributes the correlation among the constituents of the systems, ultimately entangling the optics and microwave outputs. Remarkably, obtaining maximum output entanglement does not require matching cooperativities, which is similar to what we obtained previously for frequency conversion~\cite{engelhardtOptimalBroadbandFrequency2022} in the same two-stage setup, however with stricter limitations imposed by temperature and port-coupling efficiencies.

We benchmark the entanglement capacity on the continuous variable teleportation scheme by Vaidman, Braunstein and Kimble (VBK)~\cite{vaidmanTeleportationQuantumStates1994, braunsteinTeleportationContinuousQuantum1998}: the state to be teleported is mixed with one of the parts (which we label ``a'' and represents the output optical mode) of an entangled bipartite state. The result is homodyne measured, and the outcome is used to displace the other part of the bipartite state (which we label ``c'' and represents the output microwave mode). In the optimal case, the state of ``c'' reproduces fully the state to be teleported. The performance of this protocol depends only on the generated steady-state entanglement~\cite{adessoEquivalenceEntanglementOptimal2005} and does not require a tailored sequence of time-dependent coupling rates for performing state transfer. Finally, we propose an experimentally realizable setup (see Fig.~\ref{Fig:Intro}), consisting of a microdisk resonator coupled to a microwave resonator and an optical fiber. Our material of choice throughout the manuscript is yttrium iron garnet (YIG), a staple choice in magnonics experiments that provides long lifetime of magnon excitations and large spin density~\cite{mallmannYttriumIronGarnet2013,klinglerGilbertDampingMagnetostatic2017,maier-flaigTemperaturedependentMagneticDamping2017,kosenMicrowaveMagnonDamping2019}. Alternative materials are mentioned in the outlook. Using results from simulation and experimentally feasible estimates we predict a maximum teleportation fidelity of about $0.75$.

The manuscript is structured as follows. In Section~\ref{sec:model} we introduce the model for the optomagnomechanical system~\cite{engelhardtOptimalBroadbandFrequency2022} and outline the blue-detuned optical driving scheme. Relevant parameters are summarized in Table~\ref{tab:values}. In Section~\ref{sec:outputent} we quantify the entanglement between the output microwave and optical modes using the logarithmic negativity~\cite{adessoQuantificationScalingMultipartite2004}. Afterwards in Section~\ref{sec:teleportsc} we introduce the VBK continuous variable teleportation scheme~\cite{vaidmanTeleportationQuantumStates1994, braunsteinTeleportationContinuousQuantum1998}, and quantify the expected fidelity for coherent and squeezed states. We discuss imperfections that might reduce the fidelity in form of temperature and internal versus external dissipation. Finally we summarize our results and give an outlook in form of simulation results for a prospective geometry that can host our proposed system. 

\section{Model}
\label{sec:model}
\begin{figure}[ht]
    \centering
    \includegraphics[width = 0.48\textwidth]{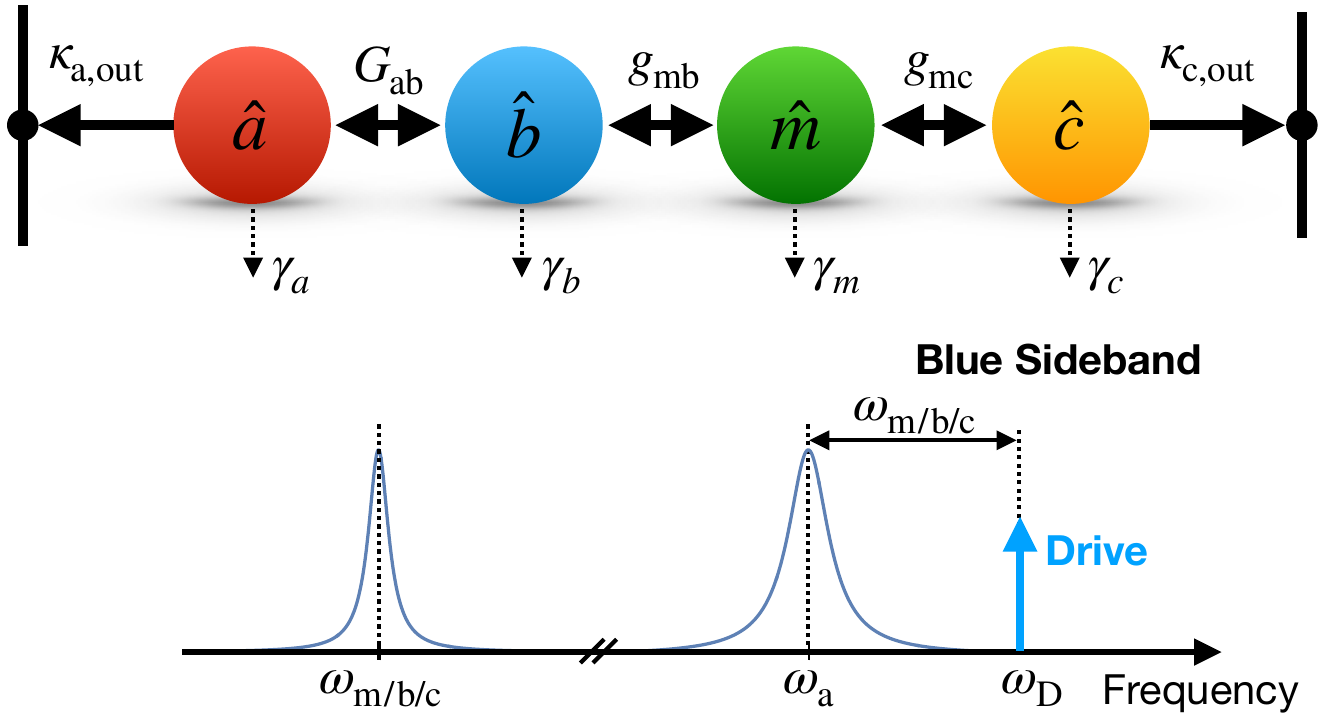}
    \caption{Schematic of the coupled system under consideration. As in the Hamiltonian Eq.~\eqref{eq:Hamiltonian1}, we denote optical photons as $\hat{a}$, phonons as $\hat{b}$, magnons as $\hat{m}$, and microwave photons $\hat{c}$. All modes have internal dissipation rates $\gamma_{\rm{a,b,m,c}}$, and optical and microwave modes are additionally coupled to output ports with rates $\kappa_{\rm{a/c, out}}$. Below the coupling scheme we show the considered driving scheme. The optical drive is placed at the blue sideband of the optical mode.}
    \label{Fig:CouplingScheme}
\end{figure}

We consider a device fabricated of a magnetic material (compare Fig.~\ref{Fig:Intro}) in which an optical mode $\hat{a}$, an elastic mode $\hat{b}$, a magnon mode $\hat{m}$ and a microwave mode $\hat{c}$ are co-localized. We consider modes $\hat{b}$, $\hat{m}$ and $\hat{c}$ to be close in frequency, while the optical mode $\hat{a}$ exhibits a higher frequency. In our envisioned device, the modes are coupled in the chain-like configuration shown in Fig.~\ref{Fig:CouplingScheme}. Photons of mode $\hat{a}$ interact with the phonons of mode $\hat{b}$ via radiation-pressure: the elastic vibrations modulate the frequency of the optical mode~\cite{aspelmeyerCavityOptomechanics2014a}. Phonons couple via the magnetoelastic interaction~\cite{spencerMagnetoacousticResonanceYttrium1958,brownMagnetoelasticInteractions1966} to magnons in mode $\hat{m}$ and, in contrast to existing works \cite{fanMicrowaveOpticsEntanglementCavity2023,liuNonreciprocalMicrowaveOptical2025}, we work with a resonant interaction between magnons and phonons. Finally, microwave modes $\hat{c}$ couple via the standard magnetic-dipole interaction to magnons~\cite{hueblHighCooperativityCoupled2013,pottsStrongMagnonPhoton2020}. A more detailed modeling of each of those interactions and the corresponding coupling rates can be found in~\cite{engelhardtOptimalBroadbandFrequency2022}. Typically, the optomagnonic and electromechanical couplings can be neglected in comparison with the other coupling rates.
The dynamics of the different modes of such a device is modeled by the bosonic Hamiltonian
\begin{equation}
    \label{eq:Hamiltonian0}
    \begin{aligned}
    \frac{\hat{H}}{\hbar} &= \omega_a \hat{a}^\dagger \hat{a} + \omega_m \hat{m}^\dagger \hat{m} + \omega_b \hat{b}^\dagger \hat{b} + \omega_c \hat{c}^\dagger \hat{c} \\
    &+ g_{ab} \hat{a}^\dagger \hat{a}(\hat{b}^\dagger +\hat{b}) + g_{mb} (\hat{b}^\dagger + \hat{b})(\hat{m}^\dagger + \hat{m}) \\
    &+ g_{mc}(\hat{m}^\dagger + \hat{m})(\hat{c}^\dagger + \hat{c}) + \nicefrac{\hat{H}_{\rm{Drive}}}{\hbar}\,, 
    \end{aligned}
\end{equation}
where $\hat{H}_{\rm{Drive}}=i\xi\left(\hat{a}e^{i\omega_Dt}-\hat{a}^{\dagger}e^{-i\omega_Dt}\right)$ is a coherent drive of the optical mode with amplitude $\xi$. Under a strong drive, we can focus on the dynamics of the weak fluctuations $\hat{a} = \alpha + \delta \hat{a}$, where $\alpha$ is a coherent amplitude of the optical mode. Keeping only second order terms in the fluctuations $\delta \hat{a}$ yields the quadratic Hamiltonian (written in a frame rotating with $\omega_D \hat{a}^{\dagger}\hat{a}$)
\begin{equation}
    \label{eq:Hamiltonian1}
    \begin{aligned}
    \frac{\hat{H}}{\hbar} &= -\Delta \delta\hat{a}^\dagger \delta\hat{a} + \omega_m \hat{m}^\dagger \hat{m} + \omega_b \hat{b}^\dagger \hat{b} + \omega_c \hat{c}^\dagger \hat{c} \\
    &+ G (\delta\hat{a}^\dagger  + \delta\hat{a})\left(\hat{b}^{\dagger}+\hat{b}\right) + g_{mb} (\hat{m}^\dagger \hat{b} + \hat{m} \hat{b}^\dagger ) \\ &+ g_{mc}(\hat{m}^\dagger \hat{c} + \hat{m} \hat{c}^\dagger)\,,
    \end{aligned}
\end{equation}
where the detuning is given by $\Delta = \omega_D - \omega_a$. Due to the spatial confinement of the optical photons, the optomechanical coupling $G=\alpha g_{ab}$ scales with the square-root of the steady-state coherent number of photons inside the cavity $\alpha = \sqrt{\bar{n}_a}$ (see Appendix~\ref{app:langevin} Eq.~\eqref{eq:meanphotonnumber}). We applied a rotating wave approximation (RWA), neglecting the terms proportional to $\hat{b}^{\dagger}\hat{m}^{\dagger}$ and $\hat{m}^{(\dagger)}\hat{c}^{(\dagger)}$. Also from now on, we adopt the notation $\delta \hat{a} \rightarrow \hat{a}$. 

In the quadratic Hamiltonian of Eq.~\eqref{eq:Hamiltonian1}, modes $\hat{b}$, $\hat{c}$ and $\hat{m}$ are close in resonance and can, depending on the coupling rates, form hybrid modes $\hat{d}_i$ with frequencies $\omega_{i}$ (with $i=1,2,3$). The optical detuning $\Delta$ can then be used to activate different interactions between optical photons and the hybrid modes. Taking the parameters summarized in Tab. \ref{tab:values}, $g_{\rm{mc}}\gg g_{\rm{mb}}$, so that the hybrid modes are given by $\hat{d}_{1/2}=\nicefrac{1}{\sqrt{2}}\left(\hat{m}\pm\hat{c}\right)$ at $\omega_{1/2} = \omega_{\rm{m/c}}\pm g_{\rm{mc}}$ and $\hat{d}_3=\hat{b}$ at $\omega_b$. For a red detuning $\Delta \approx -\omega_b$, meaning that one drives at a frequency that is lower than the cavity resonance frequency by $\omega_b$, the relevant resonance is a beam-splitter interaction $\hat{a}^\dagger \hat{d}_i +h.c.$, which allows energy exchange between optics and the other modes by combining phonons and optical drive photons into resonant cavity photons. This is the basis for optics-to-microwave frequency conversion proposed in Ref.~\cite{engelhardtOptimalBroadbandFrequency2022}. Here, we instead focus on the blue optical detuning case, $\Delta \approx \omega_b$, for which the relevant resonance is  pair-creation and annihilation $\hat{a}^\dagger \hat{d}_i^\dagger + h.c.$, since the drive frequency is now larger than the cavity resonance frequency by $\omega_b$. This interaction induces two-mode squeezing, which generates entanglement~\cite{wallsQuantumOptics2025}. Our proposal harnesses such a correlation to entangle the output of the optical and the microwave modes, which can then propagate to other systems and be used as a resource for quantum information protocols, such as state transfer. Quantifying the capacity of our system for generating entanglement requires solving for the dynamics of the coupled equations of motions describing each mode.\\

We describe the open dynamics of the system via the Heisenberg-Langevin equations for each mode $\hat{o} = \hat{a},\hat{b}, \hat{m}, \hat{c}$
\begin{equation}
   \dot{\hat{o}} = - \frac{i}{\hbar} [\hat{o}, \hat{H}] - \frac{\kappa_{\rm{o, tot}}}{2} \hat{o} - \sqrt{\kappa_{\rm{o, out}}} \hat{o}_{\rm{in}}\,,
\end{equation}
where $\kappa_{\rm{o, tot}} = \gamma_{\rm{o}} + \kappa_{\rm{o, out}}$ is the total decay rate of mode $\hat{o}$, composed of an intrinsic contribution $\gamma_{\rm{o}}$ (e.g. optical absorption for the optical mode) and a contribution due to coupling to any output port $\kappa_{\rm{o, out}}$. The noise operators $\hat{o}_{\rm{in}}$ include both intrinsic and input noise contributions, and are assumed to be include thermal and vacuum fluctuations, given by
\begin{align}
    \langle\hat{o}_{\rm{in}}\left(t\right)\hat{o}_{\rm{in}}^{\dagger}\left(t^{\prime}\right)\rangle &= \left(n_{\rm{th},o}+1\right)\delta\left(t-t^{\prime}\right),\nonumber\\
    \langle\hat{o}_{\rm{in}}^{\dagger}\left(t\right)\hat{o}_{\rm{in}}\left(t^{\prime}\right)\rangle &= n_{\rm{th},o}\delta\left(t-t^{\prime}\right)\,.
    \label{Eq:NoiseCorrelatorsBare}
\end{align}
The Heisenberg-Langevin equations (See Appendix~\ref{app:langevin} Eqs.~\eqref{eq:protocol_langevin_a} -~\eqref{eq:protocol_langevin_d} for the explicit form) can be put in convenient matrix form
\begin{equation}
    \dot{\hat{\textbf{A}}}=\mathbb{A} \hat{\textbf{A}} + \mathbb{B} \hat{\textbf{A}}_{\textrm{in}} \,,
    \label{eq:Langevin_bare}
\end{equation}
where the vector $\hat{\textbf{A}}$ contains all creation and annihilation operators, and $\hat{\textbf{A}}_{\textrm{in}}$ all the noise operators. Furthermore, the matrix $\mathbb{A}$ contains the elements of the equations of motion and matrix $\mathbb{B}$ the external and internal dissipation rates. More details can be found in Appendix~\ref{app:langevin}. 
\section{Output entanglement}
\label{sec:outputent}
As outlined in the previous section, the two-mode squeezing interaction that is activated when the optical mode is driven at blue-detuning generates entangled pairs of excitations. Combined with beam-splitter interactions between the phonons, magnons and microwave photons (in the form of Fig. \ref{Fig:CouplingScheme}) the entanglement is distributed among all the constituents of the setup and eventually generates an entangled multipartite steady-state. In particular, the optical and microwave modes can become entangled, which we refer to intra-cavity entanglement. The modes $\hat{a}$ and $\hat{c}$ couple to output ports, and thus the output modes, obtained via the standard input-output relation (for $\hat{o} = \hat{a},\hat{c}$)
\begin{equation}
    \hat{o}_{\rm{out}} = \hat{o}_{\rm{in}} + \sqrt{\kappa_{o,{\rm{out}}}} \hat{o}\,,
\end{equation}
also get entangled. Here we are interested in the the entanglement between the output modes, since they are the ones accessible experimentally and they can be used for performing state teleportation between optical and microwave photons. In this case, a state encoded in the input of the optics (microwave) mode is fully teleported to the output of the microwave (optics) mode, which can then propagate to other devices.
\subsection{Covariance matrix and the negativity}
Since the dynamics of the system, described by the Heisenberg-Langevin equations Eq.~\eqref{eq:Langevin_bare}, are linear and we assume initial states of the modes to be vacuum, the steady-state of the system is Gaussian. Therefore, the state is fully characterized by its covariance matrix, containing the first and second moments of the modes' quadratures. We define 
\begin{equation}
    \hat{\textbf{R}}=\left(\hat{x}_a, \hat{p}_a, \hat{x}_b, \hat{p}_b, \hat{x}_m, \hat{p}_m, \hat{x}_c, \hat{p}_c\right)\,,
\end{equation} 
where, e.g., $\hat{x}_a=\hat{a}^{\dagger}+\hat{a}$ and $\hat{p}_a=i\left(\hat{a}^{\dagger}-\hat{a}\right)$. 
The components of the covariance matrix are then given in terms of the correlations of the elements of $ \hat{\textbf{R}}$ defined by
\begin{equation}
    \sigma_{ij}=\frac12 \langle \hat{R}_i\hat{R}_j + \hat{R}_j\hat{R}_i\rangle - \langle \hat{R}_i \rangle \langle \hat{R}_j \rangle\,.
\end{equation}
The state of two selected subsystems, e.g. optical ($\hat{a}$) and microwave photons ($\hat{c}$), can be obtained by discarding the columns and rows of $\sigma$ corresponding to the traced-out modes ($\hat{b}$ and $\hat{m}$).

With the covariance matrix describing the joint optics-microwave state, we can compute entanglement with the logarithmic negativity. Such a quantifier is based on the Simon-Duan criterion~\cite{simonPeresHorodeckiSeparabilityCriterion2000, duanInseparabilityCriterionContinuous2000}: separable states do not produce negative symplectic eigenvalues under partial transposition. Hence for entangled states, these negative symplectic eigenvalues can quantify 'how far' a quantum state is from being separable. This is the continous variable version of the postivie-partial-transpose (PPT) or Peres-Horodecki criterion~\cite{peresSeparabilityCriterionDensity1996,horodeckiSeparabilityCriterionInseparable1997}. Given the covariance matrix between two modes $i$ and $j$ in the form:
\begin{equation}
\sigma_{ij}=
    \begin{pmatrix}
    B & C \\
    C^T & B^{\prime} 
    \label{eq:bipartiteSigmaij}
\end{pmatrix} \,,
\end{equation}
one can define the logarithmic negativity $E_N$ \cite{vidalComputableMeasureEntanglement2002} as
\begin{equation}
    E_N = \textrm{max}\left[0, -\textrm{ln}( \eta_-)\right]\, ,
    \label{eq:lognegEN}
\end{equation}
with $\eta_-$ the negative symplectic eigenvalue of $\sigma_{ij}$ under partial transposition with respect to one of the modes, and it is given by
\begin{equation}
    \eta_- = \frac{1}{\sqrt{2}} \sqrt{\Sigma - \sqrt{\Sigma^2 - 4 \,\textrm{Det}(\sigma_{ij})}}\,,
    \label{eq:log_negativity}
\end{equation}
where $\Sigma = \textrm{Det}B + \textrm{Det}B^{\prime} - 2\,\textrm{Det}C$.
\subsection{Output entanglement}
The output covariance matrix of our system is given by (see Appendix~\ref{app:covmatfrqdomain} for the full derivation)
\begin{equation} \label{eq:covmatfrequencydomain}
    \sigma^{\rm{output}}[\omega]= \frac12 \left(S_{\textbf{R}}[\omega]N_{\textbf{R}} S_{\textbf{R}}^T[-\omega]+S_{\textbf{R}}[-\omega]N_{\textbf{R}}^TS_{\textbf{R}}^T[\omega]\right),
\end{equation}
%
where $S_{\textbf{R}}$ is the scattering matrix and $N_{\textbf{R}}$ contains the quadrature noise elements. To assess the entanglement between optical and microwave outputs, we have to introduce filters at appropriate center-frequencies~\cite{sahuEntanglingMicrowavesLight2023,wangBipartiteTripartiteOutput2015}. For example, in the frame rotating at the cavity drive frequency, the optical mode is given via
\begin{equation}
    \hat{A}_{\rm{out}}\left(\omega,\sigma\right) = \int_{-\infty}^{\infty}  d\omega^{\prime } \, f\left(\omega - \omega^{\prime}, \sigma\right) \hat{a}_{\rm{out}}\left(\omega^{\prime }\right)\,,
    \label{eq:filteredmodes}
\end{equation}
where $\sigma$ is the bandwidth of the applied filter function $f$. Filtering the output modes assures that only photons at appropriate frequencies are compared when quantifying the entanglement. The filter bandwidth must be considered because in reality one only has access to a finite time interval when taking measurements of a given mode, and therefore to a finite spectral interval. From now on we refer to $\omega$ as the filter center frequency, and consider exclusively the case of a narrow bandwidth filter. We quantify the effect of a finite filter bandwidth in Appendix~\ref{app:covmatfrqdomain}. 
\begin{figure}[ht]
    \centering
    \includegraphics[width = 0.48\textwidth]{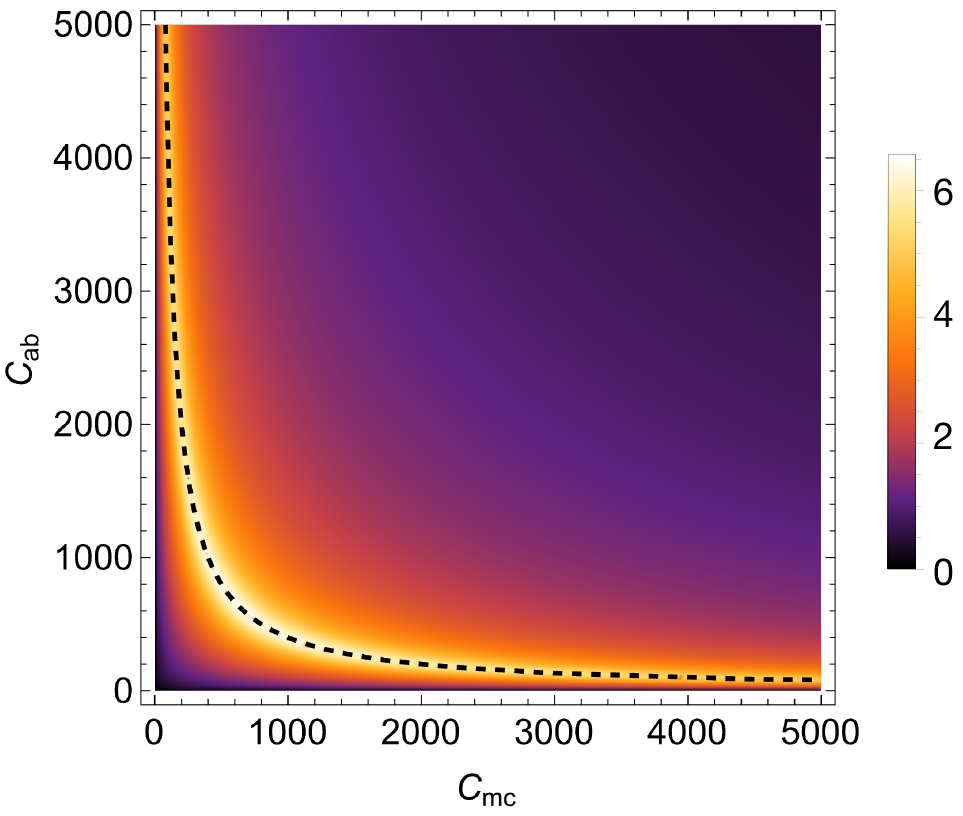}
    \caption{Logarithmic negativity $E_N$ as function of optomechanical ($C_{\rm{ab}}$) and magnon-microwave ($C_{\rm{mc}}$) cooperativity. The magnomechanical cooperativity was set to $C_{\rm{mb}}=4\times10^5$ which is based on to the values in Tab.~\ref{tab:values} and ensures comparability with Ref.~\cite{engelhardtOptimalBroadbandFrequency2022}. The black dashed line corresponds to setting the denominator equal to zero in Eq.~\eqref{eq:covmatelecoop} of Appendix~\ref{app:covmatfrqdomain}, given by the function $C_{\rm{ab}}=1+\nicefrac{C_{\rm{mb}}}{(C_{\rm{mc}}+1)}$, and corresponds to an unstable parameter regime.}
    \label{Fig:LogNegC}
\end{figure}
For the case that $\omega_m=\omega_b=\omega_c=\Delta_a$ and considering all  modes are filtered according to Eq.~\eqref{eq:filteredmodes} with center frequency $\omega = \Delta_a$, the output covariance matrix has the form 
\begin{equation}
    \sigma^{\rm{output}}_{\rm{ac}}=
    \begin{pmatrix}
        c_1 & 0 & 0 & c_3\\
        0 & c_1 & c_3 & 0\\
        0 & c_3 & c_2 & 0\\
        c_3 & 0 & 0 & c_2
    \end{pmatrix} \,,
    \label{eq:covmatinC}
\end{equation}
where $c_{1,2,3}$ are functions of the cooperativities $C_{ij}=4g_{ij}^2 /\gamma_{i,\rm{tot}} \gamma_{j,\rm{tot}}$. The exact functional forms are not illuminating and  are provided in Appendix~\ref{app:covmatfrqdomain}. The choice of the filter center frequency is pivotal, since it assures that modes are considered at appropriate frequencies in the entanglement quantification, namely an optical photon at $\omega_a$ and a microwave photon at $\omega_c$. The logarithmic negativity is calculated using Eq.~\eqref{eq:lognegEN}. In typical magnomechanical systems at the microscale, it is possible to bring selected magnon and phonon modes in resonance by tuning an external magnetic field, whereas the magnon-phonon coupling is fixed by the magnet's geometry which determines the allowed modes and their overlap. Thus, the main controllable parameters that define the steady-state of the outputs are the optomechanical and the magnon-microwave cooperativities, $C_{ab}$ and $C_{mc}$ respectively. The optomechanical cooperativity $C_{ab}$ can be set by the input power applied to the optical mode, while $C_{mc}$ can be changed by placement of the magnet in a microwave cavity (in the case of a 3D cavity), or by changing the relative position of the magnet with respect to a planar cavity.

In Fig.~\ref{Fig:LogNegC} we plot the logarithmic negativity $E_{\rm{N}}$ of the state shown in Eq.~\eqref{eq:covmatinC} as a function of the optomechanical and magnon-microwave cooperativity at fixed magnomechanical cooperativity. Using the same parameters (see Tab.~\ref{tab:values}), we notice an identical qualitative behavior as for the maximum conversion efficiency discussed in Fig. $3a)$ of Ref.~\cite{engelhardtOptimalBroadbandFrequency2022}. In particular, one does not require matching cooperativities to reach maximum $E_{\rm{N}}$, as it would be the case for one-stage setups. 
Physically, this corresponds to a point where the system reaches a parametric instability. This is special for the case of blue optical detuning, since it drives a particle non-conserving process. So when the rate at which pairs of excitations are created is not compensated by the system's dissipation rates, the linear system never reaches a steady-state. For red optical detuning, as in the case for frequency conversion, only particle conserving processes are relevant, and the system is instead pushed close to a transition to different mode hybridisation. 
\begin{figure}[ht]
    \centering
    \includegraphics[width = 0.48\textwidth]{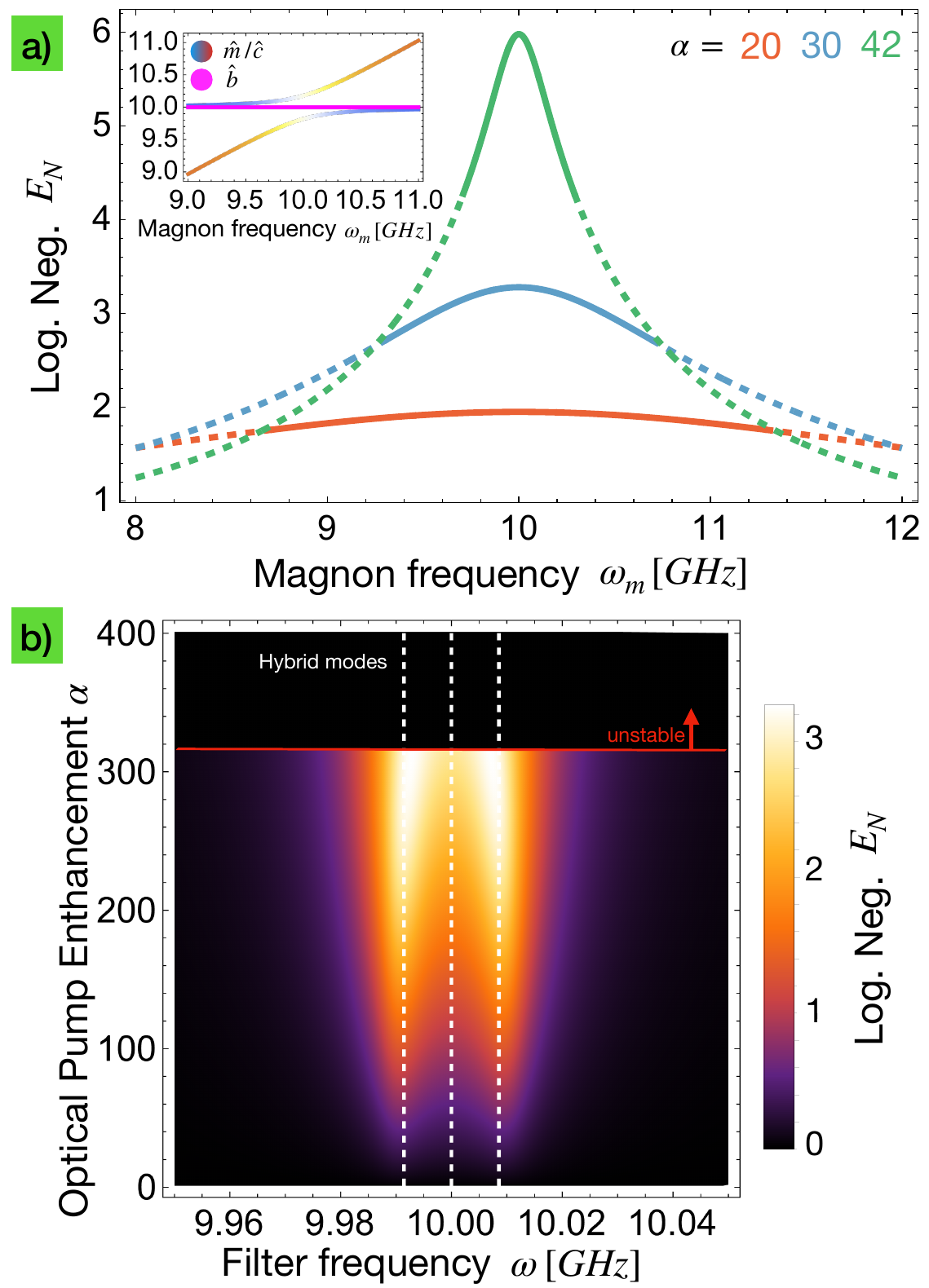}
    \caption{a) Output entanglement quantified in terms of the logarithmic negativity as a function of the magnon frequency $\omega_m$. We vary the optomechanical pump enhancement factor $\alpha$, all other parameters are in accordance with Tab.~\ref{tab:values}. Dotted lines indicate where the system becomes unstable. The inset shows the mode hybridisation as a function magnon frequency for the phonon-magnon-microwaves subsystem. In plot b), we reduce the magnon-microwave coupling to $g_{\rm{mc}}=18\,\rm{MHz}$, and plot the logarithmic negativity both as a function of the filter center frequency $\omega$, and the enhancement factor $\alpha$. The reduced coupling leads to a shift of the entanglement maxima in accordance with the changed hybrid mode frequencies, indicated by the white dashed vertical lines. The red horizontal line signifies the transition to the unstable regime.}   
    \label{Fig:InstabAnalysis}
\end{figure}
This is illustrated in Fig.~\ref{Fig:InstabAnalysis}a) by plotting $E_{\rm{N}}$ as a function of the magnon frequency and for different optomechanical coupling rates $G$, which define the strength of the pair-creation interaction and can be increased by the optical pump power. As $G$ increases, so does $E_{N}$, while simultaneously the region of stability becomes more narrow. The largest $G$ ($\alpha\approx42$) puts the system in the maximum entanglement region, in agreement with Fig.~\ref{Fig:LogNegC}. In Fig.~\ref{Fig:InstabAnalysis}b) we reduce the magnon-microwave coupling rate by one order of magnitude, which results in a shift of the regions of largest entanglement generation in accordance with the change of hybridization in the system. Hybrid mode frequencies can be easily obtained by numerically calculating the eigenvalues of matrix $\mathbb{A}$ in Eq.~\eqref{eq:Langevin_bare}~\cite{engelhardtOptimalBroadbandFrequency2022}. It is important to note that we have so far considered zero temperature, and no internal dissipation of the optical and microwave modes, hence the port coupling efficiencies are $1$. Under these circumstances we can conclude that classical frequency conversion and quantum bipartite entanglement can be optimized for the same parameters. The effect of internal losses together with other imperfections will be discussed in section~\ref{ssec:teleportfid}.\\
Experimentally, the elements of the covariance matrix can be reconstructed via homodyne measurements. This is done by mixing each output with a strong local coherent signal in a 50:50 beam-splitter, and then measuring the photocurrent in one of the beam-splitter ports. The experimental signal consists of time-dependent optical and microwave heterodyne currents, which fluctuate in time and in each experimental run. Those signals can be used to calculate correlations in Fourier domain, which, when averaged over all experimental runs, can be directly related to antisymmetrized correlations~\cite{wisemanQuantumMeasurementControl2009}, which are then used to reconstruct the covariance matrix. This is a standard procedure for measuring gaussian correlations used in different state-of-the-art experiments, e.g. in~\cite{sahuEntanglingMicrowavesLight2023}. Measurements via heterodyning are also possible, with the advantage of simpler setups but the drawback of more added noise \cite{scullyQuantumOptics1997}.
\section{Teleportation State-Transfer}
\label{sec:teleportsc}
\subsection{VBK Setup}
In this section we evaluate the performance of the system considering a particular task which harnesses output entanglement. Entanglement is a required resource to perform teleportation~\cite{streltsovColloquiumQuantumCoherence2017}. For continuous variable systems, the teleportation protocol proposed by Vaidman, Braunstein and Kimble (VBK)~\cite{vaidmanTeleportationQuantumStates1994, braunsteinTeleportationContinuousQuantum1998} uses two-mode squeezed states, and displacements conditioned on homodyne measurements. The goal of the scheme is to transfer a continuous variable state from a party, which we label ``$a$'', to another party, which we label ``$c$''. Party ``$a$'' is assumed to share a two-mode squeezed state with ``$c$''. The state to be transferred is mixed via a 50:50 beam-splitter with ''$a$'''s part of the two-mode squeezed state. Balanced homodyne detection is performed, yielding the complex amplitude $z$. Such an amplitude determines a displacement $\hat{D}\left(z\right)$ to be applied to the  ``$c$" part of the two-mode squeezed state. Such a protocol uses steady-state entanglement to improve the transfer fidelity for states of continuous variables~\cite{adessoEquivalenceEntanglementOptimal2005}. In our setup, the output entanglement is generated by the two-mode squeezing operation provided by the blue-driven optomechanical interaction.

\begin{figure*}
    \centering
    \includegraphics[width = 0.75\textwidth]{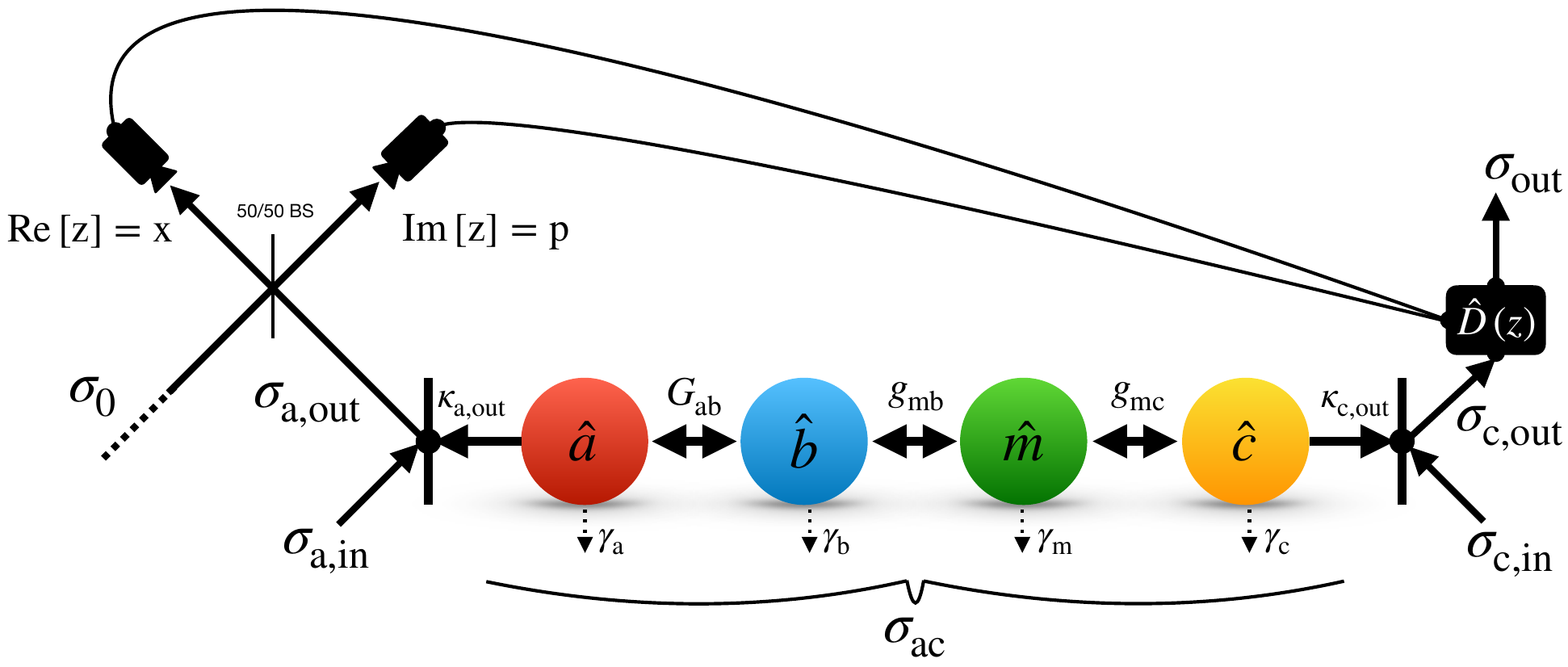}
    \caption{Sketch of the VBK transfer scheme~\cite{vaidmanTeleportationQuantumStates1994, braunsteinTeleportationContinuousQuantum1998}. The coupled chain, containing optical photons ($\hat{a}$), phonons ($\hat{b}$), magnons ($\hat{m}$) and microwaves ($\hat{c}$) generates a two-mode optics-microwave squeezed state $\sigma_{\rm{ac}}$. Teleportation of a state $\sigma_{0}$ consists on performing a displacement in the output microwave $\sigma_{\rm{out}}$ conditioned on the result of a homodyne detection in the optical part, indicated by the amplitude $z$.}
    \label{Fig:Scheme_Teleportation}
\end{figure*}
Therefore, the two-mode squeezed state of the original VBK setup is obtained as the entangled output from the hybrid chain as depicted in Fig.~\ref{Fig:Scheme_Teleportation}. At zero temperature, the performance of this protocol is only dependent on the amount of steady-state entanglement that can be generated between input and output. This is set by the system design and the optical drive, and therefore does not require the control of the intermediate coupling rates. Such a protocol is better suited to magnomechanical systems compared to state-transfer protocols that require temporal control of coupling strengths or to use dark modes to mitigate dissipation loss~\cite{wangUsingDarkModes2012}.
\subsection{Teleportation fidelity} \label{ssec:teleportfid}
\begin{figure}[h!]
    \centering
    \includegraphics[width = 0.48\textwidth]{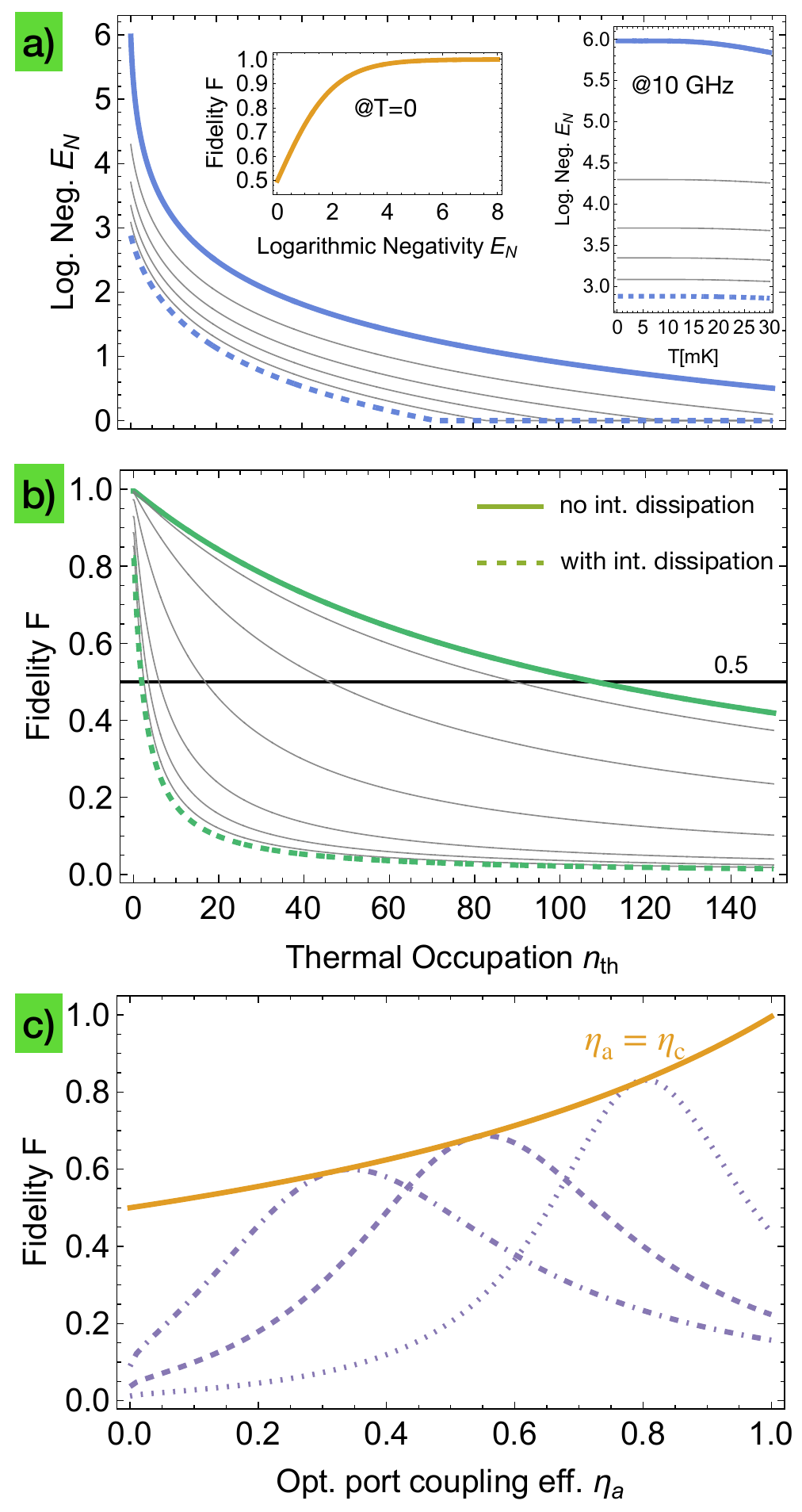}
    \caption{Logarithmic negativity a) and teleportation fidelity b) of an coherent state as a function of temperature, without (solid) and with (dashed) internal mode dissipation of optics and microwave, set to $\gamma_{\rm{a, int}} = 1\, \rm{GHz}$ and $\gamma_{\rm{c, int}} = 1\, \rm{MHz}$, respectively. The left inlet in  plot a) shows the fidelity of teleporting a coherent state as a function of logarithmic negativity at zero temperature. The right inlet shows low-temperature entanglement results for $-\Delta=\omega_{\rm{b/m/c}}=10\,\rm{GHz}$. The gray plot lines in a) and b) indicate increasing values of the internal dissipation rates $\gamma_{\rm{a/c, int}}$ from $0$ to the values mentioned above. c) State teleportation fidelity for a coherent state as a function of the optical port coupling efficiency $\eta_{\rm{a}}$. It is obtained by fixing the total linewidths $\gamma_{\rm{a/c, tot}} = \gamma_{\rm{a/c, int}}+\kappa_{\rm{a/c, out}} = 1.1\, \rm{GHz}$/100.1\, \rm{MHz} and putting $\alpha=42$. The purple lines corresponds to different fixed values of the microwave port coupling efficiency $\eta_{\rm{c}} = 0.35, 0.55, 0.8$. The orange line is obtained by setting $\eta_{\rm{a}} = \eta_{\rm{c}}$.}
    \label{Fig:TempEffect}
\end{figure}
For the teleportation of a coherent Gaussian state $\sigma_0$, the teleportation fidelity of the VBK scheme can be calculated (see Appendix~\ref{app:fidelity} for a detailed derivation) as
\begin{align}
    F = \frac{2}{ \sqrt{\det\left(\sigma_{0}\right)\det\left(\sigma_{\rm{out}}\right) \det \left( \sigma_{0}+\sigma_{\rm{out}}\right)}}\,.
    \label{eq:FidelityFinalExpression}
\end{align}
Besides depending on the state to be teleported $\sigma_0$, the fidelity is characterized by the covariance matrix $\sigma_{\rm{out}}$ (compare Eq.~\eqref{eq:WignerOutputCompact}) of the output of the magnomechanical system, which, in turn, depends on the amount of entanglement generated. Thermal noise is most detrimental to the entanglement generation, as it causes destruction of the delicate quantum states. In contrast to the optical two-mode squeezed state in the original VBK proposal, our system is especially prone to thermal effects, since three of the four modes operate in the microwave frequency regime where thermal occupation at temperatures above zero is expected to cause significant decoherence. In the following we quantify the generated output steady-state entanglement as a function of temperature with a special emphasis on the ratio of internal to external dissipation rates at the two ports of our setup.\\

In Fig.~\ref{Fig:TempEffect} a) and b) we plot both the output logarithmic negativity, and the teleportation fidelity for a coherent state, as a function of thermal occupation. We chose this state because it is the Gaussian state with minimal uncertainty and is of special interest for applications in quantum optics and quantum information. For the case of no internal dissipation of optical and microwave modes, we observe that at around $n_{\rm{th}}\approx 105$ (corresponding to about $50.6\,\rm{K}$ at $10\,\rm{GHz}$) the teleportation fidelity surpasses $0.5$ for the optimum case with no internal dissipation of microwave and optical photon modes. This point signifies an improvement over the best fully classical local approach~\cite{ferraroGaussianStatesContinuous2005}. When internal dissipation is considered, entanglement starts to be generated at lower thermal occupations, and the fidelity is increased in a even smaller temperature region (on the order of a few $100\,\rm{mK}$ at $10\,\rm{GHz}$). This suggests that internal mode dissipation at the port sites leads to much reduced teleportation fidelity even when entanglement is generated, an effect linked to the impact of port coupling efficiency.\\ 

In Fig.~\ref{Fig:TempEffect} c) we plot the teleportation fidelity as a function of the optical port coupling efficiency for three different values of the microwave port coupling efficiency at zero temperature. We observe a rapid decline of fidelity when either one of the port coupling efficiencies deviates from being equal to the other. For example, the port coupling efficiencies for the dashed curve in Fig.~\ref{Fig:TempEffect} b) are $\eta_{\rm{a}} = 0.91$ and $\eta_{\rm{c}} = 0.99$ which reduces the maximally achievable fidelity to about $0.82$. Furthermore, Fig.~\ref{Fig:TempEffect}c) reveals that the teleportation fidelity can be optimized for reduced port coupling efficiencies, as long as they are matching. This implies a more complicated dependency of the entanglement generation on the port coupling efficiencies compared to the classical frequency conversion efficiency, which is just linearly dependent on $\eta_{\rm{a,c}}$~\cite{engelhardtOptimalBroadbandFrequency2022}. The results also depend on the respective total linewidths of optical and microwave photon modes, which in turn are subject to the conditions on the cooperativities for maximizing entanglement discussed in the previous section. In Fig.~\ref{Fig:Benchmark_TeleportationFidelity} of Appendix~\ref{app:fidelity} we further show the fidelity for squeezed states as a function of the logarithmic negativity. Our results suggest that in our setup, teleportation of squeezed states can be achieved at a reduced fidelity at a given $E_{N}$.
Regarding setup imperfections, temperature and internal dissipation rates are the most detrimental to the achievable entanglement. 
\section{Implementation in a magnomechanical disk} \label{sec:implementation}
\begin{table}[h]
\caption{\label{tab:valuesdisk}Values for mode frequencies, linewidths/dissipations and coupling strengths for the proposed disk geometry.}
\begin{ruledtabular}
\begin{tabular}{ccc}
 Quantity  &Symbol &Value \\ \hline
  Optical photon freq.                & $\nicefrac{\omega_a}{2\pi}$           & $193.5\,\rm{THz}$  \\ 
 
 Optical detuning (blue)               & $\nicefrac{\Delta_a}{2\pi}$           & $0.567\,\rm{GHz}$  \\
 
 Phonon freq.                        & $\nicefrac{\omega_b}{2\pi}$           & $0.567\,\rm{GHz}$  \\
 
 Magnon freq.                       & $\nicefrac{\omega_m}{2\pi}$           & $0.567\,\rm{GHz}$  \\

 Microwave freq.                       & $\nicefrac{\omega_c}{2\pi}$           & $0.567\,\rm{GHz}$  \\

 Internal optical linewidth         & $\nicefrac{\gamma_a}{2\pi}$           & $0.001\,\rm{THz}$  \\
 
 External optical dissipation       & $\nicefrac{\kappa_{a,\rm{out}}}{2\pi}$     & $0.099\,\rm{THz}$  \\
 
 Phonon linewidth                   & $\nicefrac{\gamma_b}{2\pi}$           & $5\,\rm{kHz}$  \\
 
 Magnon linewidth                   & $\nicefrac{\gamma_m}{2\pi}$           & $5\,\rm{MHz}$  \\
 
 Internal mw cavity loss rate      & $\nicefrac{\gamma_c}{2\pi}$            & $0.01\,\rm{MHz}$  \\
 
 External mw cavity coupling rate     & $\nicefrac{\kappa_{c,\rm{out}}}{2\pi}$     & $0.99\,\rm{MHz}$  \\
 
 Magnomechanical coupling           & $\nicefrac{g_{mb}}{2\pi}$             & $5\,\rm{MHz}$  \\
 
 Magnon-Microwave photon coupling          & $\nicefrac{g_{mc}}{2\pi}$      & $10\,\rm{MHz}$  \\

 Optomechanical coupling            & $\nicefrac{g_{ab}}{2\pi}$             & $1.0\,\rm{kHz}$ \\
 
 Optomechanical coop. (base value)            & $C_{ab}$             & $8\times10^{-9}$ \\

 Magnomechanical coop.            & $C_{mb}$             & $4000$ \\

 Magon-microwave coop.            & $C_{mc}$             & $80$ \\
\end{tabular}
\end{ruledtabular}
\end{table}
We now assess the implementation of the state teleportation scheme in a YIG disk, a promising geometry for tailoring the resonant magnon-phonon coupling~ \cite{bondarenkoMagnetoelasticConversionIntegrated2026}. For this proposed geometry, a realistic set of parameters is given in Table~\ref{tab:valuesdisk}, based on simulations and calculations presented in Appendix~\ref{app:estimates}. We consider a YIG disk with radius $3.7\, \mu\rm{m}$, thickness $200\,\rm{nm}$ and with a hole of $400\,\rm{nm}$ mounted on a stem, see Fig.~\ref{Fig:Intro}. The disk hosts optical whispering gallery modes (WGMs), which can couple to the evanescent light field of an optical fiber placed near its edge, that serves as an optical output port. The last component of the device is a microwave ring resonator, whose microwave field couples to the magnetic excitations of the disk.

Compared to Ref.~\cite{bondarenkoMagnetoelasticConversionIntegrated2026}, we choose a two-times thicker disk, which improves the optomechanical interaction strength, our main source of entanglement. Based on our numerics (see Fig.~\ref{Fig:app_numericaldetails} c) of Appendix \ref{app:estimates}), we expect an optimized single-photon optomechanical coupling rate of about $1\,\rm{kHz}$, for the only azimuthal telecom wavelength WGM with mode number $m=20$ and a radially symmetric mechanical breathing mode, which have been identified as the optimal modes for the protocol~\cite{bondarenkoMagnetoelasticConversionIntegrated2026}. The increased thickness slightly lowers the magnomechanical coupling rate compared to the optimal result, to $5\,\rm{MHz}$ for a magnetic vortex breathing mode. In terms of the inductive magnon-microwave coupling, a mode volume overlap of $0.01\%$ already suffices to obtain a coupling rate on the order of $10\,\rm{MHz}$~\cite{engelhardtOptimalBroadbandFrequency2022}. Based on estimates given in Appendix \ref{app:estimates}, see Eq.~\eqref{eq:magnonmicrowavecoup}, we choose the coupling rate to be $8\,\rm{MHz}$. For the bare optical mode we assume $1550\,\rm{nm}$ telecom-wavelength photons, corresponding to a frequency $193.5\,\rm{THz}$. For the protocol the cavity should  be driven blue detuned by the frequency of the phonon mode. The mechanical mode frequency is fixed by the  the geometry and sets the necessary frequencies for all other modes in the system, including the optical detuning. The magnon mode frequency is tunable by an external magnetic field. Our numerical analysis performed as in Ref.~\cite{bondarenkoMagnetoelasticConversionIntegrated2026} indicates that for this geometry and given dimensions, we can set all modes to resonance at $0.567\,\rm{GHz}$ for an optimized working point.

\begin{figure}[h!]
    \centering
    \includegraphics[width = 0.48\textwidth]{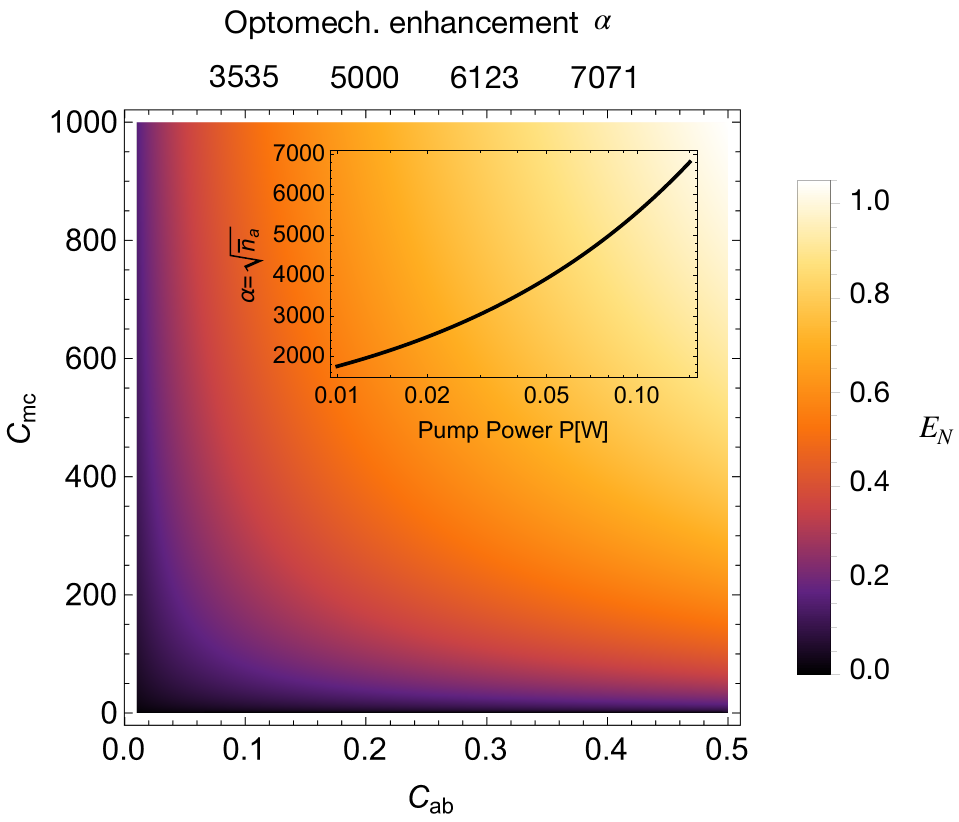}
    \caption{Steady state entanglement as a function of optomechanical $(C_{\rm{ab}})$ and magnon-microwave $(C_{\rm{mc}})$ cooperativity, for the proposed disk-based setup. The magnomechanical cooperativity was set, based on numerical analysis, to $C_{\rm{mb}} = 4000$. The inlet shows the enhancement factor $\alpha$ as a function of the pump power. The upper scale shows the factor $\alpha$ for three values of $C_{\rm{ab}}$. The values of the relevant parameters are given in Table~\ref{tab:valuesdisk}.}   
    \label{Fig:EstimateAnalysis}
\end{figure}
Concerning the dissipative behavior of the modes, we resort again to numerical analysis and to estimates based on the literature. For optical modes at telecom frequency, our numerically obtained Q-factor is approximately $2700$, corresponding to a total linewidth of about $0.1\,\rm{THz}$ (see Fig.~\ref{Fig:app_numericaldetails} b) of Appendix \ref{app:estimates}). This puts the setup outside of the resolved sideband regime, since $\omega_b<\gamma_{\rm{a,tot}}$. In this regime, the interaction that generates entanglement, a two-mode squeezing interaction between the optical mode and the phonon mode, cannot be addressed perfectly, which furthers reduces the achievable entanglement. We assume a split of this total linewidth into an intrinsic part of the cavity mode itself (e.g. due to absorption), and an external part to the output optical fiber as $\gamma_{\rm{a,\,tot}} = \gamma_{\rm{a}} + \kappa_{\rm{a,\,ext}} = 0.001\,\rm{THz} + 0.099\,\rm{THz}$, which can be justified based on the very low optical absorption of YIG in the telecom wavelength regime~\cite{boothMagnetoopticPropertiesRare1984,vojnaVerdetConstantMagnetoActive2019}. This overcoupling of the optical port ensures optimal results for the entanglement generation, as has been discussed in the previous section. Based only on these numbers, we have $C_{\rm{ab}} \approx 10^{-8}$, thus a very strong optical drive is required to enable significant entanglement generation. The dissipation of the magnon mode is determined by the Gilbert damping, and is expected for YIG microstructures to be on the order of $5\,\rm{MHz}$ \cite{anCoherentLongrangeTransfer2020}. Quality factors for mechanical modes in YIG disks are yet to be measured, however based on published results in similar silicon based geometries \cite{anetsbergerUltralowdissipationOptomechanicalResonators2008,nguyenUltrahighQfrequencyProduct2013a}, we assume a mechanical linewidth on the order of $5\,\rm{kHz}$, yielding $C_{\rm{mb}} = 4000$. 
Lastly, for the microwave resonator we assume a total linewidth $\gamma_{\rm{c,\,tot}} = 1\,\rm{MHz}$ with a split between intrinsic and external as $\gamma_{\rm{c}} + \kappa_{\rm{c,\,ext}} = 0.01\,\rm{MHz} + 0.99\,\rm{MHz}$. Planar ring resonators can exhibit high internal quality factors~\cite{sunSuperconductingRingResonators2025}, which allows the operation of the resonator in the overcoupling regime via proper designing its coupling to the output ports. This numerical values yield the magnon-microwave cooperativity $C_{\rm{mc}}=80$, which can be tuned further in experiment, for example by adjusting the distance between resonator and disk, which yields a larger coupling due to better mode overlap. 

Given the above considerations, we display in Fig.~\ref{Fig:EstimateAnalysis} the result for steady-state logarithmic negativity as function of both the optomechanical and magnon-microwave cooperativity. The primer can be controlled by the optical drive power, while the latter is mostly determined by mode overlap set by the distance between the split-ring resonator and the magnetic disk. Even though the optomechanical cooperativity is very low due to the small expected single-photon coupling, a logarithmic negativity of $E_N\approx1$ can be achieved within a reasonable parameter space. Such an amount of entanglement corresponds to a teleportation fidelity of $F\approx0.75$, as benchmarked in Fig.~\ref{Fig:Benchmark_TeleportationFidelity}.

\section{Conclusions and Outlook}
%
%
%
We investigated the output entanglement between optical and microwave modes using a resonantly coupled magnomechanical hybrid system as mediator. This systems incorporates resonant magnon-phonon and magnon-microwave couplings with the radiation pressure interaction between optical photons and phonons. The entanglement generated by a blue-detuned driven optomechanical interaction is maximized by pushing the system towards instability, with maximum entanglement obtained without the requirement of matching different cooperativities. This allows for a flexibility in the system's design not present in other platforms. Interestingly, such a parameter regime coincides with the optimal frequency conversion regime obtained in a previous work~\cite{engelhardtOptimalBroadbandFrequency2022}.

The entanglement between output microwaves and optical photons was benchmarked by assessing the system's performance for continuous variable state teleportation. We have considered the scheme by Vaidman, Braunstein and Kimble, which has the major advantage of not relying on any temporal control of the couplings of dark-state generation. Our analysis predicts that unit fidelities can be achieved for transferring coherent states at dilution fridge temperatures. Coherent state teleportation sets an important benchmark for potential advantages of using entangled states for quantum information tasks \cite{ferraroGaussianStatesContinuous2005}, such as teleportation of non-Gaussian states. Here, an important aspect is the impact of port-coupling efficiency which ought to be matched for optimal results. Besides the system design, such a scheme requires only standard tools, such as homodyne measurements and displacement operations, that are available both in the optical and in the microwave regime.

Finally, we proposed a device based on a mounted ferrimagnetic YIG-disk geometry \cite{valetFieldTheoryLinear2025, valetOrbitalAngularMomentum2025, bondarenkoMagnetoelasticConversionIntegrated2026} where our protocol can be implemented. Using both numerical simulations and estimates for coupling strengths and linewidths, we predict a fidelity of up to $0.75$ for coherent state teleportation. The corresponding maximum output entanglement is on the order of $E_{N}\approx 1$, which outperforms the current experimentally achieved value of $E_N\approx0.17$~\cite{sahuEntanglingMicrowavesLight2023}. Our analysis assumes noiseless operation of all the measurement apparatuses, which is an idealization. Under real conditions, added noise in the measurement chain would further reduced the amount of useful entanglement. This could be mitigated by exploring more efficient designs, such as microstructured crystals, which already have been fabricated and probed~\cite{rashediPhotonicCrystalCavities2025}. Furthermore, assuming a large mechanical Q-factor ensures entanglement generation at reasonable optical pump powers, even though the base-value of the optomechanical cooperativity remains small. Once such devices are properly optimized, we expect that its optomechanical coupling can reach that of similar devices fabricated with silicon \cite{qiuLaserCoolingNanomechanical2020}, yielding results that are much closer to our predicted theoretical optimum.

Our proposed protocol represents a timely application of magnomechanical interaction for quantum information. The unique tunability of the magnons together with their strong resonant interaction with phonons in optimized geometries renders magnomechanical hybrid systems a promising platform for mediating the large frequency gap between optical and microwave frequencies, facilitating core tasks of quantum information processing. Recent experiments~\cite{mullerChiralPhononsPhononic2024, mullerTemperatureDependenceMagnonphonon2024} performed on bilayer structures at temperatures in the single Kelvin range further show the growing interest of operating magnomechanical systems in the quantum regime, or more precisely, in a regime where entanglement can be generated. A natural extension of this work would be to generalize the discussion to non-Gaussian states, such as superposition states. Even though calculations are equivalent for these states, analytical expressions for e.g. the fidelity are unlikely to be obtained and a numerical analysis is needed. Finally, the large generated output entanglement also goes alongside symmetric steerability (Appendix~\ref{app:steering}), which guarantees efficient distribution of entanglement among the constituents of our setup and enables implementation of other information protocols such quantum key distribution.
\section*{Acknowledgements}
We thank Stefano Chesi, Sanchar Sharma, Benjamin Pigeau and Olivier Klein for very useful discussions. This work was supported by the EU-project HORIZON-EIC-2021-PATHFINDER OPEN PALANTIRI-101046630. F. E. and S.V.K. also acknowledge partial financial support by the Federal Ministry of Research, Technology and Space (BMFTR) project QECHQS (Grant No. 16KIS1590K). 
\appendix
\begin{table}[h!]
\caption{\label{tab:values}Values for mode frequencies, linewidths/dissipations and coupling strengths taken from Ref.~\cite{engelhardtOptimalBroadbandFrequency2022}.}
\begin{ruledtabular}
\begin{tabular}{ccc}
 Quantity  &Symbol &Value \\ \hline
  Optical photon freq.                & $\nicefrac{\omega_a}{2\pi}$           & $200\,\rm{THz}$  \\ 
 
 Optical detuning (blue)               & $\nicefrac{\Delta_a}{2\pi}$           & $10\,\rm{GHz}$  \\
 
 Phonon freq.                        & $\nicefrac{\omega_b}{2\pi}$           & $10\,\rm{GHz}$  \\
 
 Magnon freq.                       & $\nicefrac{\omega_m}{2\pi}$           & $10\,\rm{GHz}$  \\

 Internal optical linewidth         & $\nicefrac{\gamma_a}{2\pi}$           & $0.1\,\rm{GHz}$  \\
 
 External optical dissipation       & $\nicefrac{\kappa_{a,\rm{out}}}{2\pi}$     & $1\,\rm{GHz}$  \\
 
 Phonon linewidth                   & $\nicefrac{\gamma_b}{2\pi}$           & $1\,\rm{kHz}$  \\
 
 Magnon linewidth                   & $\nicefrac{\gamma_m}{2\pi}$           & $1\,\rm{MHz}$  \\
 
 Internal mw cavity loss rate      & $\nicefrac{\gamma_c}{2\pi}$            & $1\,\rm{MHz}$  \\
 
 External mw cavity coupling rate     & $\nicefrac{\kappa_{c,\rm{out}}}{2\pi}$     & $100\,\rm{MHz}$  \\
 
 Magnomechanical coupling           & $\nicefrac{g_{mb}}{2\pi}$             & $10\,\rm{MHz}$  \\
 
 Magnon-Microwave photon coupling          & $\nicefrac{g_{mc}}{2\pi}$      & $180\,\rm{MHz}$  \\

 Optomechanical coupling            & $\nicefrac{g_{ab}}{2\pi}$             & $0.2\,\rm{MHz}$ \\
\end{tabular}
\end{ruledtabular}
\end{table}
\section{Langevin equations} \label{app:langevin}
\begin{figure}[h!]
    \centering
    \includegraphics[width = 0.48\textwidth]{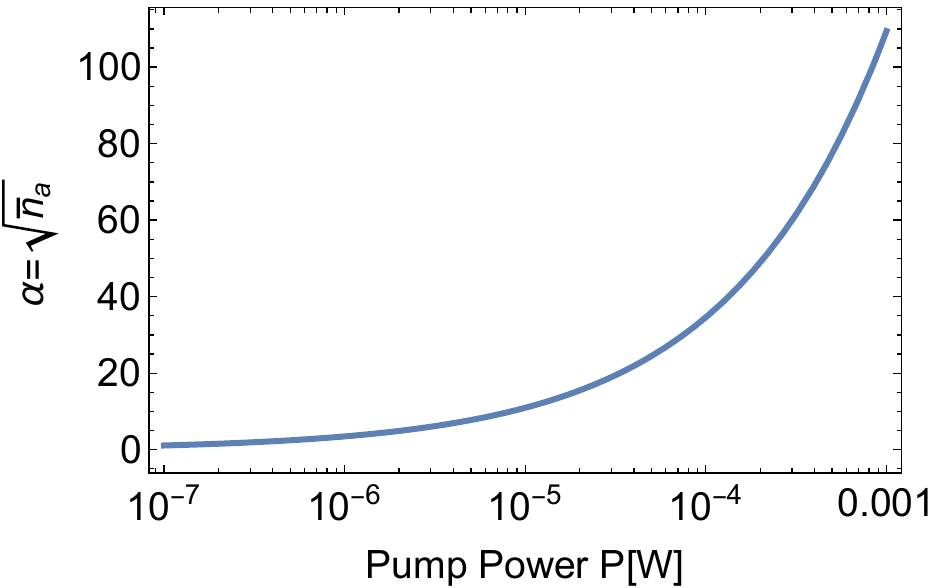}
    \caption{Plot of the square root of Eq.~\eqref{eq:meanphotonnumber} as a function of the optical pump power using parameters from Tab.~\ref{tab:values}.}
    \label{Fig:PumpEnhacement}
\end{figure}

The parameters used in the main text are summarized in Table~\ref{tab:values}. They are virtually identical to the ones used in Ref.~\cite{engelhardtOptimalBroadbandFrequency2022}. The mean photon number in the optical cavity $\bar{n}_a$ can be estimated by 
\begin{equation}
    \bar{n}_a=\frac{\kappa_{a,\rm{out}}P}{\hbar \omega_a \left[\Delta_a ^2 + \left (\gamma_a + \kappa_{a,\rm{out}} \right)^2/4 \right]}\,,
    \label{eq:meanphotonnumber}
\end{equation}
where $P$ is the drive power, $\gamma_a$ the internal optical mode linewidth, $\kappa_{a,\rm{out}}$ the dissipation rate to the external port (see Fig.~\ref{Fig:PumpEnhacement}). We adopt the notation $\delta \hat{a} \rightarrow \hat{a}$.
The quantum Langevin equations (QLE's) for the bosonic annihilation operators are obtained in the usual way via the Heisenberg equation of motion $\dot{\hat{o}}\left(t\right)=-i[\hat{o},\hat{H}]/\hbar$ (for an arbitrary bosonic operator $\hat{o}$), as
\begin{subequations}
\begin{eqnarray}
    \dot{\hat{a}}\left(t\right)=
    &&i\Delta - \frac{\gamma_{a}+\kappa_{a, \rm{out}}}{2}\hat{a} -i G_{ab}\left(\hat{b}+\hat{b}^{\dagger}\right)\nonumber\\
    &&-\sqrt{\kappa_{a}}\hat{a}_{\rm{in}}\left(t\right),\label{eq:protocol_langevin_a}\\
    \dot{\hat{b}}\left(t\right)=
    && -i\omega_b - \frac{\gamma_b}{2}\hat{b} -i G_{ab}\left(\hat{a}+\hat{a}^{\dagger}\right) -i g_{mb}\hat{m}\nonumber\\
    &&-\sqrt{\gamma_{b}}\hat{b}_{\rm{in}}\left(t\right),\\
    \dot{\hat{m}}\left(t\right)=
    && -i\omega_m - \frac{\gamma_m}{2}\hat{m} -i g_{mb}\hat{b}
    -i g_{mc}\hat{c}\nonumber \\
    &&-\sqrt{\gamma_{m}}\hat{m}_{\rm{in}}\left(t\right),\\
    \dot{\hat{c}}\left(t\right)=
    &&-i\omega_c -\frac{\gamma_{c}+\kappa_{c, \rm{out}}}{2}\hat{c} -i g_{mc}\hat{m}\nonumber\\
    &&-\sqrt{\kappa_c}\hat{c}_{\rm{in}}\left(t\right)\,.
\label{eq:protocol_langevin_d}
\end{eqnarray}
\end{subequations}
These equations can be cast in the from of Eq.~\eqref{eq:Langevin_bare} in the main text by using vectors 
\begin{equation}
    \hat{\textbf{A}} = \left(\hat{a}^{\dagger},\hat{a},\hat{b}^{\dagger},\hat{b},\hat{m}^{\dagger},\hat{m},\hat{c}^{\dagger},\hat{c}\right)^{T}\,,
\end{equation}
and
\begin{align}
    &\hat{\textbf{A}}_{\rm{in}} = \\
    &\left(\hat{a}^{\dagger}_{\rm{out}},\hat{a}_{\rm{out}},\hat{a}^{\dagger}_{\rm{in}},\hat{a}_{\rm{in}},\hat{b}^{\dagger}_{\rm{in}},\hat{b}_{\rm{in}},\hat{m}_{\rm{in}}^{\dagger},\hat{m}_{\rm{in}},\hat{c}_{\rm{out}},\hat{c}_{\rm{out}}^{\dagger}, \hat{c}_{\rm{in}}^{\dagger},\hat{c}_{\rm{in}}\right)^{T}\,.
\end{align}
The transformation matrix connecting the bare mode operators $\hat{\textbf{A}}$ with the quadrature operators $\hat{\textbf{R}}$ via $\hat{\textbf{R}}=Q \hat{\textbf{A}}$. Recall that the quadratures are defined here as
\begin{equation}
    \begin{aligned}
        \hat{x}_a &= \hat{a}+\hat{a}^\dagger\,, \\
        \hat{p}_a &= i(\hat{a}^\dagger-\hat{a})\,.
    \end{aligned}
\end{equation}
hence the transformation matrix reads
\begin{equation}
Q = \mathbb{1}_6  \bigotimes 
        \begin{pmatrix}
            1 & i\\
            -i & 1
        \end{pmatrix}       
\end{equation}
%
%
%
%
\section{Covariance matrix in the frequency domain} \label{app:covmatfrqdomain}
\begin{figure}[h!]
    \centering
    \includegraphics[width = 0.48\textwidth]{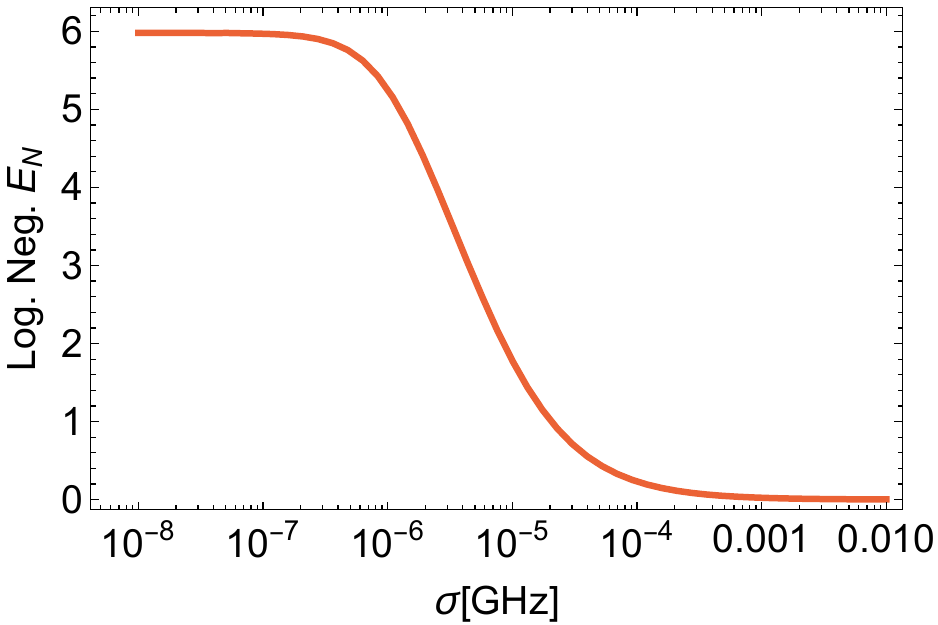}
    \caption{Generated output entanglement $E_N$ as a function of the filter bandwidth $\sigma$.} 
    \label{Fig:FilterDependency}
\end{figure}

Our aim is to obtain the covariance matrix for the optics and microwave output in frequency space. In general, the elements of the covariance matrix in frequency space are given by
\begin{equation}
    \sigma_{ij}^{\rm{intra}}[\omega]= \frac12 \int_{-\infty}^{\infty} d\omega^{\prime} \langle \{ \hat{R}_i [\omega], \hat{R}_j[\omega^{\prime}]\}\rangle \,,
\end{equation}
where $\{\}$ is the anti-commutator and where we can already assume a vacuum ground state, so $\langle \hat{R}_i\rangle\langle \hat{R}_j\rangle = 0$. In the next step we conveniently write
\begin{equation}
    \sigma^{\rm{intra}}[\omega]= \frac12 \int_{-\infty}^{\infty} d\omega^{\prime} \left\langle  \hat{\textbf{R}}[\omega]\hat{\textbf{R}}^T[\omega^{\prime}] + \left(\hat{\textbf{R}}[\omega]\hat{\textbf{R}}^T[\omega^{\prime}]\right)^T \right\rangle \,.
\end{equation}
Since we want to quantify the output entanglement, we set $\hat{\textbf{R}} = \hat{\textbf{R}}_{\rm{out}}$. The output and input quadratures are connected by a scattering-matrix via $\hat{\textbf{R}}_{\rm{out}}=S_{\textbf{R}}[\omega]\hat{\textbf{R}}_{\rm{in}}$. For the bare modes $\hat{\textbf{A}}$, the scattering matrix is obtained by input-output relations $\hat{\textbf{A}}_{\rm{out}}=\hat{\textbf{A}}_{\rm{in}} + \mathbb{B}\hat{\textbf{A}}$ leading to $S_{\textbf{A}}[\omega] = \mathbb{B}^T\left(i \omega \mathbb{1}_8 + \mathbb{A}\right)^{-1}\mathbb{B}+\mathbb{1}_{12}$, where the matrices $\mathbb{A}$ and $\mathbb{B}$ are defined in Eq.~\eqref{eq:Langevin_bare}. In the quadrature basis we then have $S_{\textbf{R}}[\omega]=QS_{\textbf{A}}[\omega]Q^{-1}$. With that we write the covariance matrix for the outputs as
\begin{align}
    \sigma^{\rm{out}}[\omega]=
    \frac12 \int_{-\infty}^{\infty} d\omega^{\prime}  S_{\textbf{R}}[\omega]\langle \hat{\textbf{R}}_{\rm{in}}[\omega]\hat{\textbf{R}}_{\rm{in}}^T[\omega^{\prime}]\rangle S_{\textbf{R}}^T[\omega^{\prime}]\nonumber \\
    + \frac12\int_{-\infty}^{\infty} d\omega^{\prime} \left( S_{\textbf{R}}[\omega]\langle \hat{\textbf{R}}_{\rm{in}}[\omega]\hat{\textbf{R}}_{\rm{in}}^T[\omega^{\prime}]\rangle S_{\textbf{R}}^T[\omega^{\prime}]\right)^T.
\end{align}
This expression contains correlators of the input noise in the form $\langle \hat{\textbf{R}}_{\rm{in}}[\omega]\hat{\textbf{R}}_{\rm{in}}^T[\omega^{\prime}]\rangle = N_{\textbf{R}}\delta\left(\omega + \omega^{\prime}\right)$. Therefore, the final expression for the output covariance matrix reads 
\begin{equation}
    \sigma^{\rm{out}}[\omega]= \frac12 \left(S_{\textbf{R}}[\omega]N_{\textbf{R}} S_{\textbf{R}}^T[-\omega]+S_{\textbf{R}}[-\omega]N_{\textbf{R}}^TS_{\textbf{R}}^T[\omega]\right)\,,
\end{equation}
where
\begin{equation}
    N_{\textbf{R}} = \bigoplus_{j= a,b, m ,c}
    \begin{pmatrix}
        2n_{\rm{th,j}}+1 & i \\
        -i & 2n_{\rm{th,j}}+1 
    \end{pmatrix}.
    \label{eq:noisematrixR}
\end{equation}
Here, the (white) noise correlations in frequency space for the bare operators were used, defined as 
\begin{eqnarray}
    \langle \hat{\beta}\left(\omega\right)\hat{\beta}^{\dagger}\left(\omega^{\prime}\right)\rangle && =\left(\rm{n_{th,\beta}}+1\right)\delta\left(\omega+\omega^{\prime}\right)\,,\nonumber\\
    \langle \hat{\beta}^{\dagger}\left(\omega\right)\hat{\beta}\left(\omega^{\prime}\right)\rangle && =n_{\rm{th,\beta}}\delta\left(\omega+\omega^{\prime}\right)\,,
\end{eqnarray}
with $\hat{\beta}=\hat{a}_{\rm{in}}, \hat{b}_{\rm{in}}, \hat{m}_{\rm{in}}, \hat{c}_{\rm{in}}$. For the filter function in Eq.~\eqref{eq:filteredmodes} we assume a Gaussian function with width $\sigma$, given by 
\begin{equation}
    f(\sigma, \omega^{\prime}-\omega) = \frac{1}{\sqrt{\sigma\sqrt{\pi}}} \exp\left[-\frac{\left(\omega^{\prime}-\omega\right)^2}{2\sigma^2}\right] \ .
\end{equation}
The dependence on the generated entanglement as a function of the filter bandwidth $\sigma$ can be seen in Fig.~\ref{Fig:FilterDependency}.
Next, the elements of the covariance matrix defined in Eq.~\eqref{eq:covmatinC} in terms of the cooperativites $C_{ij}=4g_{ij}^2 /\gamma_{i,\rm{tot}} \gamma_{j,\rm{tot}}$ are given as
\begin{widetext}
    \begin{align} \label{eq:covmatelecoop}
        c_1 &= \frac{C_{\rm{ab}}^2 \left(1+C_{\rm{mc}}\right)^2 +6C_{\rm{ab}}\left(1+C_{\rm{mc}}\right)\left(1+ C_{\rm{mb}}+C_{\rm{mc}}\right)\left(1+C_{\rm{mb}}+C_{\rm{mc}}\right)^2}{\left(1+C_{\rm{mb}}+C_{\rm{mc}}-C_{\rm{ab}}\left(1+C_{\rm{mc}}\right)\right)^2} \ , \nonumber \\
        c_2 &= \frac{C_{\rm{ab}}^2 \left(1+C_{\rm{mc}}\right)^2 + \left(1 + C_{\rm{mb}} + C_{\rm{mc}}\right)^2 -2 C_{\rm{ab}}\left(C_{\rm{mb}}-3C_{\rm{mb}}C_{\rm{mc}}+\left(1+C_{\rm{mc}}\right)^2\right)}{\left(1+C_{\rm{mb}}+C_{\rm{mc}}-C_{\rm{ab}}\left(1+C_{\rm{mc}}\right)\right)^2} \ , \nonumber\\
        c_3 &= \frac{4\sqrt{C_{\rm{ab}}C_{\rm{mb}}C_{\rm{mc}}}\left(1+C_{\rm{ab}}+C_{\rm{mb}}+C_{\rm{mc}}\left(1+C_{\rm{ab}}\right)\right)}{\left(1+C_{\rm{mb}}+C_{\rm{mc}}-C_{\rm{ab}}\left(1+C_{\rm{mc}}\right)\right)^2} \ .
    \end{align} 
\end{widetext}
Note that due to two stages between the output modes, the covariance matrix has rotated off-diagonal blocks compared to the standard TMS covariance matrix, that is obtained for one-stage setups. More generally, the output covariance matrix takes the form
\begin{equation}
    \sigma_{\rm{general}}^{\rm{ij}}=
    \begin{pmatrix}
        a_1 & 0 & a_4 & a_3\\
        0 & a_1 & a_3 & -a_4\\
        a_4 & a_3 & a_2 & 0\\
        a_3 & -a_4 & 0 & a_2
    \end{pmatrix} \ ,
    \label{eq:covmatinGeneral}
\end{equation}
where the $a_i$ are real numbers. Based on Eqs.~\eqref{eq:lognegEN} and~\eqref{eq:log_negativity} we have
\begin{equation*}
    \Sigma = a_1^2 + a_2^2 +2 \left(a_3^2 + a_4^2\right) \ ,
\end{equation*}
\begin{equation}
     \det\sigma = \left(-a_1a_2 + a_3^2 + a_4^2\right)^2 \ , 
\end{equation}
and
\begin{align}
    \eta_- = \frac{1}{\sqrt{2}} \bigg(a_1^2 + a_2^2 + 2\left(a_3^2 + a_4^2\right)\bigg. \nonumber \\ \left. - |a_1+a_2|\left(\left(a_1 - a_2\right)^2+4\left(a_3^2 + a_4^2\right)\right)^{1/2}\right)^{1/2} \ .
\end{align}
\section{Teleportation fidelity} \label{app:fidelity}
In this appendix we show the detailed derivation of the teleportation fidelity for coherent states. For the first part, we follow the first steps of Ref.~\cite{ferraroGaussianStatesContinuous2005} and then calculate an analytical expression for the output Wigner function that does not contain any further integration. This can only be done when we consider purely states with Gaussian statistics, that are fully characterized by their covariance matrix. The teleportation is then obtained by comparing the overlap of the input and output Wigner function. Note that in the following, our notation is not specific for the system in the main text. For optics-to-microwave teleportation, one would have indices $2,3 = \hat{a}, \hat{c}$ for $\textbf{x}_2$ and $\textbf{x}_3$. The Wigner function of the output state is then given by 
\begin{align}
   &W_{\rm{out}}\left(\textbf{x}_{0},\textbf{x}_3\right)=  \nonumber \\ &\int dx_z dp_z dx_2 dp_2 W\left[\sigma_{0}\right]\left(x_z + x_2, p_z - p_2\right)\nonumber \\ 
     &W\left[\sigma_{23}\right]\left(x_2, p_2, x_3 - x_z, p_3 - p_z\right) \ .
     \label{eq:Wignerinput}
\end{align}
The output Wigner function only depends on the phase space coordinates of the output state at $\textbf{x}_3$, which in our model would be, for example, the microwave output port. The integration over the coordinates $\textbf{x}_2$ and $\textbf{x}_z$ covers all possible outcomes of the joint measurement performed at the input point (in our model the optical port). The z-coordinates come from a double-homodyne density probability distribution $p(z)$, considering the kind of joint measurement that can be performed on two optical states. The Wigner function of the coherent input state (to be teleported) is given by 
\begin{align}
    W_{0}&\left(x_z + x_2, p_z - p_2\right) = \nonumber \\ &\exp\left(-\frac12 \left(\textbf{x}_2^T \sigma_z + \textbf{x}_z^T-\textbf{x}_{0}^T\right)\sigma_{0}^{-1}\right. \nonumber \\ &\left. \left(\sigma_z \textbf{x}_2 + \textbf{x}_z - \textbf{x}_{0}\right) \vphantom{\frac12}\right)\frac{1}{2\pi \sqrt{\det\left(\sigma_{0}\right)}} \,,
    \label{eq:coherentWigner}
\end{align}
where we included a displacement vector $\textbf{x}_{0}$ and where $\sigma_z$ is the Pauli-z matrix. We further introduced the notation $\textbf{x}_2=\left(x_2, p_2\right)^{T}$. For the Wigner function of the entangled output-pair, we first decompose its covariance matrix $\sigma_{23}$ into four blocks via 
$\sigma_{23}^{-1}=\begin{pmatrix}
    \mathbb{A}_2 & \mathbb{A}_{23} \\
    \mathbb{A}_{32} & \mathbb{A}_3
\end{pmatrix}$.
The resulting Wigner function is given by
\begin{align}
    W_{23} & \left(x_2, p_2, x_3 - x_z, p_3 - p_z\right) = \nonumber \\ &\exp\left(-\frac12 \left( \textbf{x}_2^T \mathbb{A}_2\textbf{x}_2 + \textbf{x}_2^T \mathbb{A}_{23} \left(\textbf{x}_3-\textbf{x}_z\right) \right. \right. \nonumber \\ &\left. \left. \nonumber + \left(\textbf{x}_3^T-\textbf{x}_z^T\right)\mathbb{A}_{32}\textbf{x}_2 + \left(\textbf{x}_3^T-\textbf{x}_z^T\right)\mathbb{A}_3 \left(\textbf{x}_3-\textbf{x}_z\right) \right)  \vphantom{\frac12} \right) \nonumber \\
    &\frac{1}{(2\pi)^2 \sqrt{\det\left(\sigma_{23}\right)}} \,.
    \label{eq:Wignerentangled23}
\end{align}
By defining the vector $\textbf{x}_{2z}=\textbf{x}_2 \bigoplus \textbf{x}_z$ we can group the terms in Eqs.~\eqref{eq:coherentWigner} and~\eqref{eq:Wignerentangled23} conveniently together to obtain 
%
%
\begin{align}
    W_{\rm{out}}\left(\textbf{x}_{0},\textbf{x}_3\right) = 
    \frac{1}{(2\pi)^3\sqrt{\det \sigma_{0}\det{\sigma_{23}}}} \nonumber \\
    \exp\left(-\frac12 \left(\textbf{x}_3^T\mathbb{A}_3\textbf{x}_3 + \textbf{x}_{0}^T\sigma_{0}\textbf{x}_{0}\right) \right)
    \int \rm{dx_z dp_z dx_2 dp_2}\nonumber \\ \exp \left(\frac12 \left(\textbf{x}_{2z}^T \mathbb{Q}_{2z}\textbf{x}_{2z} + \left(\textbf{A}_{2z}^T+\textbf{B}_{2z}^T\right)\textbf{x}_{2z}\right) \right) \,,
    \label{eq:wignerout_before_int}
\end{align}
where we have defined the matrix
\begin{equation}
\mathbb{Q}_{2z}=
    \begin{pmatrix}
    \mathbb{A}_2 + \sigma_z \sigma_{0} \sigma_z & -\mathbb{A}_{23} + \sigma_z \sigma_{0} \\
    -\mathbb{A}_{23}+\sigma_{0} \sigma_z & \mathbb{A}_3 + \sigma_{0}
\end{pmatrix}\, ,
\end{equation}
and the two vectors
\begin{equation}
\textbf{A}_{2z}=
    \begin{pmatrix}
    \mathbb{A}_{23}\textbf{x}_3 - \sigma_z \sigma_{0} \textbf{x}_{0} \\
    -\mathbb{A}_3 \textbf{x}_3 - \sigma_{0} \textbf{x}_{0}
\end{pmatrix} \,,
\end{equation}
and
\begin{equation}
\textbf{B}_{2z}=
    \begin{pmatrix}
    \mathbb{A}_{32}^T\textbf{x}_3 - \sigma_z \sigma_{0}^T \textbf{x}_{0} \\
    -\mathbb{A}_3^T \textbf{x}_3 - \sigma_{0}^T \textbf{x}_{0}
\end{pmatrix}\,.
\end{equation}
In the next step we can carry out the integral in Eq.~\eqref{eq:wignerout_before_int} by using the identity for a multi-dimensional Gaussian integral with additional linear term. The final result is
\begin{align}
    W_{\rm{out}}\left(\textbf{x}_{0},\textbf{x}_3\right)=& \frac{1}{2\pi \sqrt{\det\left(\sigma_{0}\right) \det\left(\sigma_{23}\right)\det\left(\mathbb{Q}_{2z}\right)}} \nonumber \\
    &\exp\left(-\frac12 \left(\textbf{x}_3^T\mathbb{A}_3\textbf{x}_3 + \textbf{x}_{0}^T\sigma_{0}\textbf{x}_{0}\right) \right) \nonumber \\
    &\exp \left(\frac12 \left(\textbf{A}_{2z}+\textbf{B}_{2z}\right)\mathbb{Q}_{2z}^{-1}\left(\textbf{A}_{2z}^T+\textbf{B}_{2z}^T\right)\right) \,.
    \label{eq:outputWigner}
\end{align}
The teleportation fidelity for coherent states can be calculated via
\begin{equation}
    F=4\pi \int \rm{d}\textbf{x}_3 W_{\rm{out}}\left(\textbf{x}_{0}, \textbf{x}_3\right)W\left[\sigma_{0}\right]\left(\textbf{x}_{0},\textbf{x}_3\right) \,.
\end{equation}
Using Eqs.~\eqref{eq:coherentWigner} and~\eqref{eq:outputWigner}. In order to carry out the integration we decompose the second exponential in Eq.~\eqref{eq:outputWigner} in the form
%
%
\begin{align}
    W&_{\rm{out}}\left(\textbf{x}_{0},\textbf{x}_3\right)= \frac{1}{2\pi \sqrt{\det\left(\sigma_{0}\right) \det\left(\sigma_{23}\right)\det\left(\mathbb{Q}_{2z}\right)}} \nonumber \\
    &\exp \left(-\frac12 \left(\textbf{x}_3^T \textbf{A}\textbf{x}_3 + \textbf{x}_{0}^T\textbf{B}\textbf{x}_3 + \textbf{x}_{0}^T\textbf{C}\textbf{x}_{0}\right)\right)\,,
    \label{eq:outputWignerDecomp}
\end{align}
where we defined
\begin{align}
    \textbf{A} &= -\left(\mathbb{A}_{32}\mathbb{B}_{2}\mathbb{A}_{23} - \mathbb{A}_{32}\mathbb{B}_{23}\mathbb{A}_{3} - \mathbb{A}_{3}\mathbb{B}_{32}\mathbb{A}_{23} + \mathbb{A}_{3}\mathbb{B}_{3}\mathbb{A}_{3} \right)+\mathbb{A}_3\nonumber\,,\\
    \textbf{B} &= -2\sigma_{0} \left( -\sigma_z\mathbb{B}_{2}\mathbb{A}_{23} - \mathbb{B}_{32}\mathbb{A}_{23} + \sigma_z\mathbb{B}_{23}\mathbb{A}_{23} + \mathbb{B}_{3}\mathbb{A}_{3} \right)\,,\nonumber\\
    \textbf{C} &= -\sigma_{0} \left(\sigma_z\mathbb{B}_2\sigma_z + \sigma_z\mathbb{B}_{23} + \mathbb{B}_{32}\sigma_z + \mathbb{B}_3 \right) \sigma_{0} + \sigma_{0}\,,
\end{align}
and used the elements of the matrix 
$\mathbb{Q}_{\rm{2z}}^{-1}=\begin{pmatrix}
    \mathbb{B}_2 & \mathbb{B}_{23} \\
    \mathbb{B}_{32} & \mathbb{B}_3
\end{pmatrix}$\,.
There exists an orthogonal transformation $\textbf{U}$ ($\textbf{U}^{-1}=\textbf{U}^{T}$) such that $\textbf{U}\textbf{B}=-2\textbf{A}$ and $\textbf{U}\textbf{C}\textbf{U}^{-1}=\textbf{A}$. This transforms the exponent in Eq.~\eqref{eq:outputWignerDecomp} via 
\begin{align}
    \textbf{x}_{0}^T \textbf{B}  \textbf{x}_{3} &= \textbf{x}_{0}^T \textbf{U}^{-1} \textbf{U}\textbf{B}  \textbf{x}_{0} = -\textbf{y}_{0}^T \textbf{A} \textbf{x}_3 -\textbf{x}_{3}^T \textbf{A}  \textbf{y}_{0} \,,\nonumber \\
    \textbf{x}_{0}^T \textbf{B}  \textbf{x}_{0} &=  \textbf{x}_{0}^T \textbf{U}^{-1} \textbf{U}\textbf{C} \textbf{U}^{-1} \textbf{U} \textbf{x}_{0} = \textbf{y}_{0}^T \textbf{A}  \textbf{y}_{0}\,,
    \label{eq:OrthogTrafo}
\end{align}
and the fidelity integral becomes
\begin{align}
    F &= \int \rm{d}\left(\textbf{x}_3-\textbf{y}_{0}\right) \frac{4\pi}{(2\pi)^2 \det\left(\sigma_{0}\right)\sqrt{ \det\left(\sigma_{23}\right)\det\left(\mathbb{Q}_{2z}\right)}} \nonumber \\
    &\exp \left(-\frac12 \left(\left(\textbf{x}_3-\textbf{y}_{0}\right)^T \left(\textbf{A}+\sigma_{0}^{-1}\right)\left(\textbf{x}_3-\textbf{y}_{0}\right) \right)\right)\,,
    \label{eq:FidelityThirdStep}
\end{align}
which yields the result 
\begin{align}
    F = \frac{2}{ \det\left(\sigma_{0}\right)\sqrt{ \det\left(\sigma_{23}\right)\det\left(\mathbb{Q}_{2z}\right)\det\left(\textbf{A}^{-1}+\sigma_{0}\right)}}\,.
    \label{eq:FidelityFinalStep}
\end{align}
Identifying $\det\left(\sigma_{0}\right) \det\left(\sigma_{23}\right)\det\left(\mathbb{Q}_{2z}\right) = \det\left(\sigma_{\mathrm{out}}\right)$ and $\sigma_{\rm{out}}^{-1} = \textbf{A}$ gives Eq.\eqref{eq:FidelityFinalExpression} of the main text. Notice that all coherent states have the same teleportation fidelity, since the quantity depends only on the covariance matrix of the state to be transfered $\omega_0$ and the amount of entanglement shared between the two modes used for the protocol. One can also recast the Wigner function of the output mode as
\begin{align}
    W&_{\rm{out}} \left(\textbf{y}_{0},\textbf{x}_3\right) = \nonumber \\ &\exp\left(-\frac12 \left(\textbf{x}_3-\textbf{y}_{0}\right)^T\sigma_{\rm{out}}^{-1}\left(\textbf{x}_3 - \textbf{y}_{0}\right) \vphantom{\frac12}\right)\frac{1}{2\pi \sqrt{\det\left(\sigma_{\rm{out}}\right)}} \,.
    \label{eq:WignerOutputCompact}
\end{align}
Note that the input vector transforms via $\textbf{y}_{0} = \textbf{U}\textbf{x}_{0}$ with $\textbf{U} = -2\textbf{A} \textbf{B}^{-1}$. That simplifies the expression for the fidelity even further, and finally gives rise to Eq.~\eqref{eq:FidelityFinalExpression} in the main text.
\begin{figure}[h!]
    \centering
    \includegraphics[width = 0.48\textwidth]{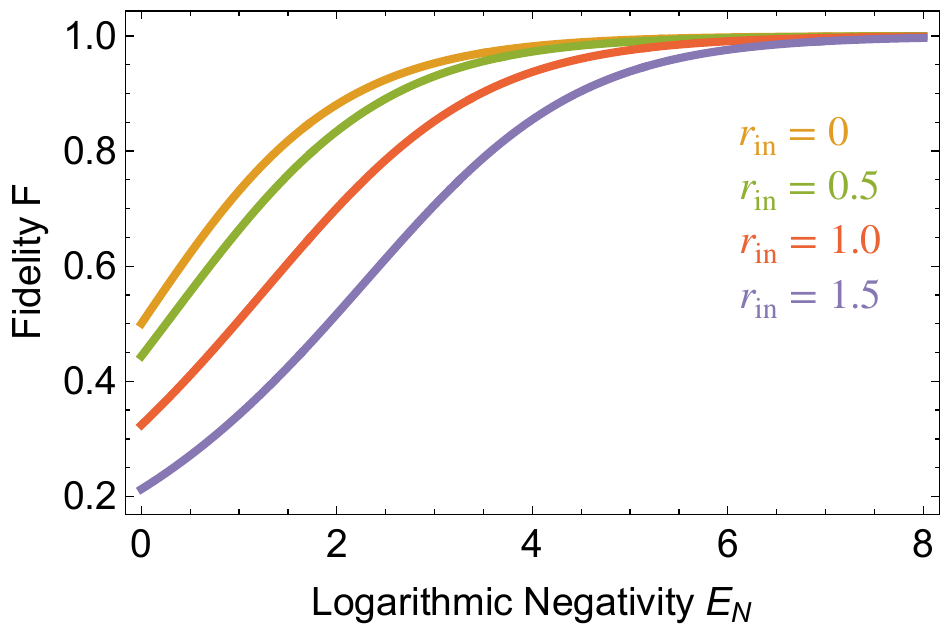}
    \caption{Teleportation fidelity as a function of logarithmic negativity calculated based on the TMS covariance matrix in Eq.~\eqref{eq:TBScovarianceMatrix}. The colors indicate different squeezing factors $r_{\rm{in}}$ of the state to be teleported.} 
    \label{Fig:Benchmark_TeleportationFidelity}
\end{figure}
For benchmarking the generated entanglement, we can compare to the covariance matrix of a twin-beam state exhibiting two-mode squeezing controlled by the parameter $r$
\begin{equation}
    \sigma_{\rm{TMS}}=
    \begin{pmatrix}
        \cosh\left(2r\right) & 0 & \sinh\left(2r\right) & 0 \\
        0 & \cosh\left(2r\right) & 0 & -\sinh\left(2r\right)\\
        \sinh\left(2r\right) & 0 & \cosh\left(2r\right) & 0 \\
        0 & -\sinh\left(2r\right) & 0 & \cosh\left(2r\right) 
    \end{pmatrix}
    \label{eq:TBScovarianceMatrix}\,.
\end{equation}
We can use equation~\eqref{eq:TBScovarianceMatrix} to benchmark the transfer fidelity as a function of the logarithmic negativity $E_{\rm{N}}$. Results for different squeezed coherent input states are plotted in Fig.~\ref{Fig:Benchmark_TeleportationFidelity}. The more squeezed the input state is, meaning the larger the squeezing factor $r_{\rm{in}}$ is chosen, the larger $E_{\rm{N}}$ of the setup that generates two-mode squeezing needs to be, in order to achieve high transfer fidelity.  
\section{Steering} \label{app:steering}
An additional quantity of interest in terms of security is entanglement steering. This quantifies how much one party, say Alice, can influence the other party, Bob, by actions on her side. For continuous-variable systems described by a bipartite covariance matrix (compare Eq.~\eqref{eq:bipartiteSigmaij}), it can be quantified as~\cite{kogiasQuantificationGaussianQuantum2015} 
\begin{equation}
    \mathbb{S}=\max\{0, \, S\left(B^{\left(\prime\right)}\right) - S\left(\sigma_{ij}\right)\}\,,
\end{equation}
where $S(\sigma)=\frac12 \ln\left(\det \sigma\right)$ is the Réniy-2 entropy function and where the use of either $B$ or $B^{\prime}$ determines the steering of $\hat{a}$ on $\hat{c}$ or vice versa, respectively. 
\begin{figure}[h!]
    \centering
    \includegraphics[width = 0.48\textwidth]{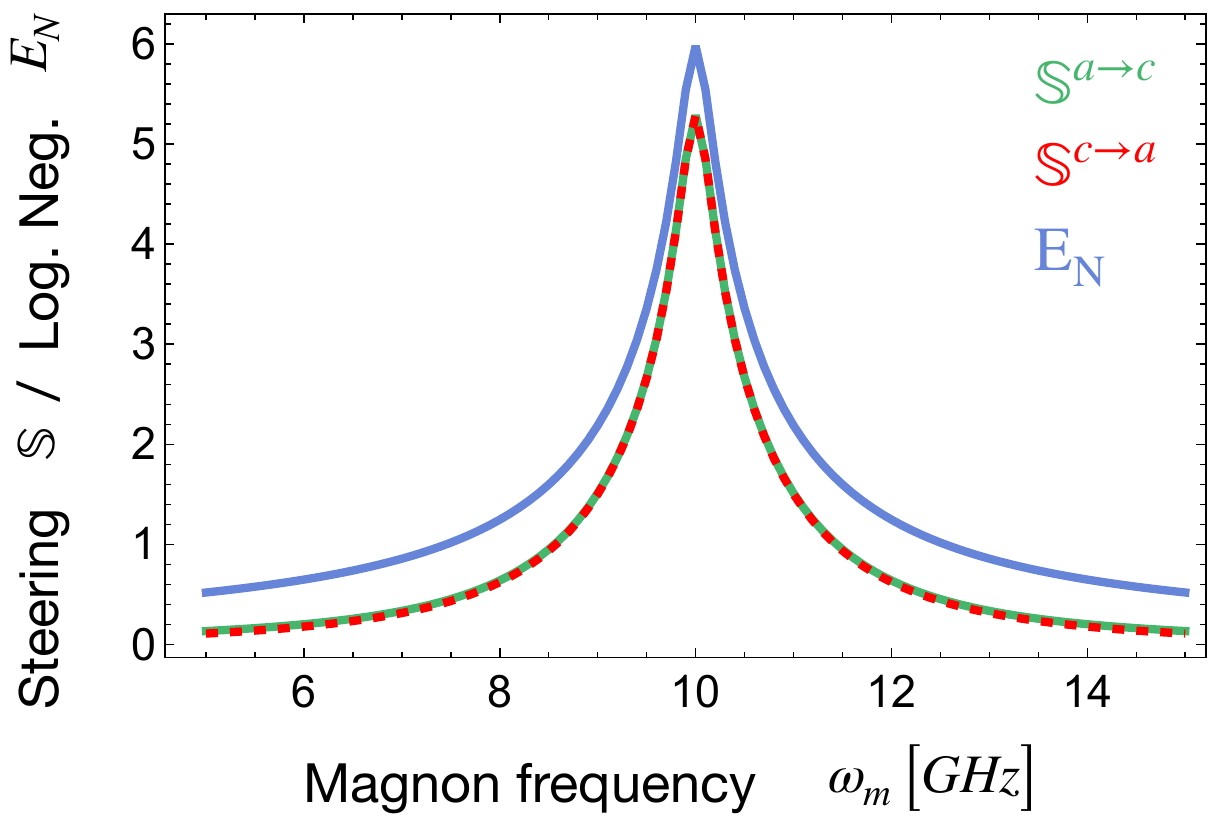}
    \caption{Quantification of quantum steering in both directions under optimal conditions alongside the generated logarithmic negativity. The temperature is set to zero, all other parameters follow Tab.~\ref{tab:values} with an optical drive enhancement of $\alpha=42$.} 
    \label{Fig:Steering}
\end{figure}
The plot in Fig.~\ref{Fig:Steering} shows the quantum steering in both directions under optimal conditions and zero temperature. We observe exactly what was reported in Ref.~\cite{kogiasQuantificationGaussianQuantum2015}. The steering is symmetric for the two-mode coherent state of our system and quantitatively always smaller than the logarithmic negativity. The maximum difference between the two quantities is about $\ln2$.  
\section{Details on Numerical Results and Estimates} \label{app:estimates}
\begin{figure*}
    \centering
    \includegraphics[width = 0.95\textwidth]{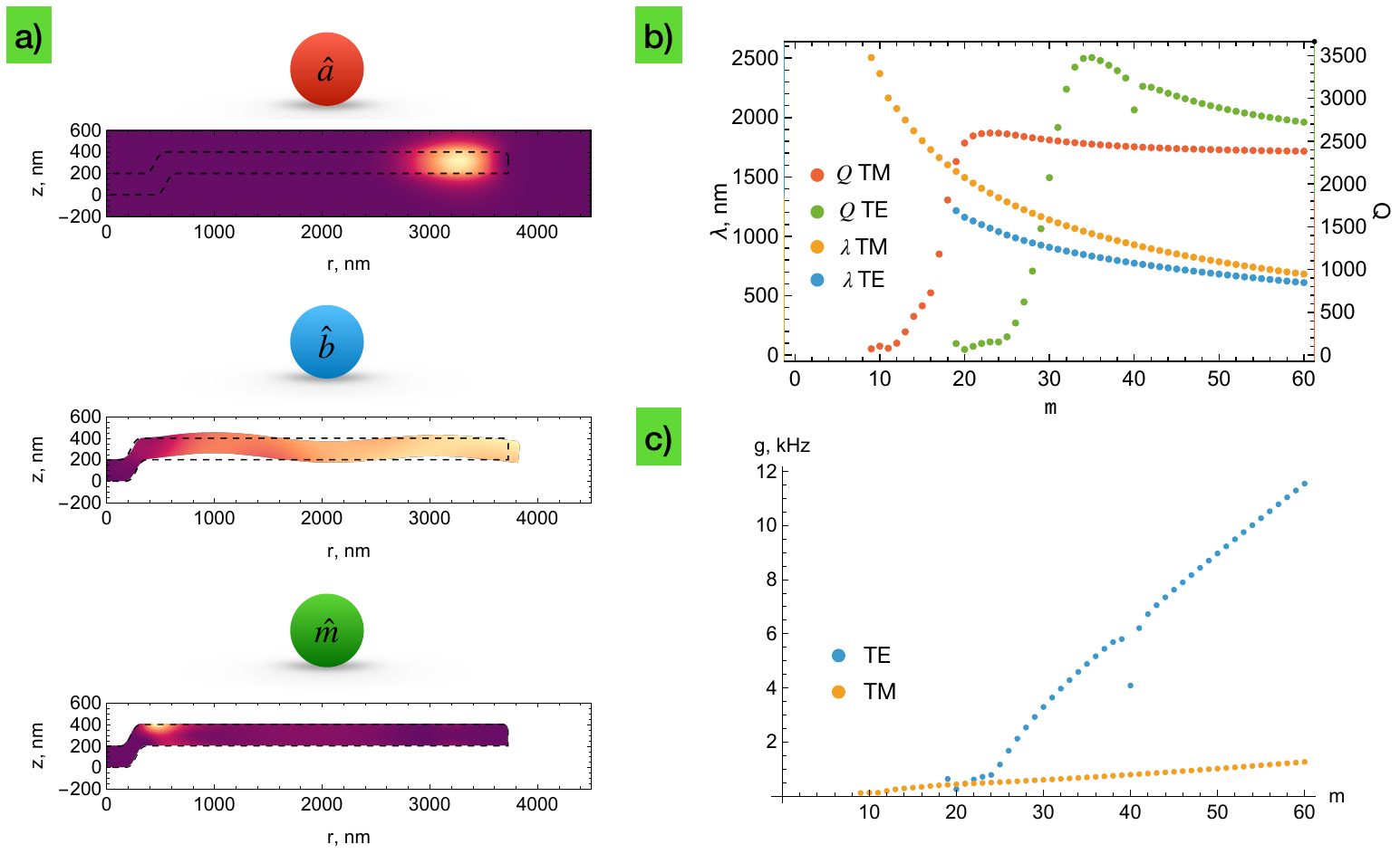}
    \caption{a) Examples of spatial mode profiles for optical photon $\hat{a}$ (azimuthal WGM, $m=20$ for telecom wavelength of $\approx1550\,\mathrm{nm}$), phonon $\hat{b}$ (radially symmetric breathing mode) and magnon $\hat{m}$ (vortex breathing mode). b) Optical mode wavelength $\lambda$ and quality factor $Q$ as a function of the azimuthal mode number $m$ for both TE and TM. c) Optomechanical coupling rate for TE and TM modes as a function of the azimuthal mode number $m$.}
    \label{Fig:app_numericaldetails}
\end{figure*}
In the following we give further details on the numerically obtained parameters used in section~\ref{sec:implementation}. Fig.~\ref{Fig:app_numericaldetails} contains the results that have been obtained based on numerical calculations. Fig.~\ref{Fig:app_numericaldetails} a) shows examples of spatial mode profiles that can be used, while b) and c) show results for the optical modes and quality factors, as well as the expected optomechanical coupling.  
\subsection{Microwave Ring Resonator}
The microwave photons are considered to be contained in a LC circuit based resonator, which is described by a Hamiltonian of the form
\begin{equation}
\hat{H}_{\rm{LC}}=\hbar\omega_{c}\left(\hat{c}^{\dagger}\hat{c}+\frac{1}{2}\right). \label{eq:app_LCcircuit}
\end{equation}
We quantize the interaction with the LC circuit using the charge-flux canonically conjugate variables. The interaction Hamiltonian is constructed solely based on the Zeeman interaction,
\begin{equation}
\label{eq:MWCoup}
    H_{\rm{Zeeman}} = \int \rm{dr}^3 \,M\left( \bf{r}\right) \cdot B\left(\bf{r}\right)\,,
\end{equation}
where the field of the coupling antenna is assumed to come from a ring of negligible thickness and width going around the nanostructure ($\approx100\,\rm{nm}$) off of its side and at the height of the foot of the structure. The required field $B$ can be found via the Biot-Savart law
\begin{equation}
    \textbf{B}\left(\textbf{r}\right) = \frac{\mu_0}{4\pi}\int_{U} \frac{I\left[dl \times \textbf{r}^{\prime} \right]}{|\textbf{r}^{\prime}|^3}\,,
\end{equation}
where we integrate over a closed curve $U$. The current can be obtained in quantized form from the charge zero-point fluctuations and the Hamiltonian~\eqref{eq:app_LCcircuit} 
\begin{equation}
    \hat{I} = \frac{\rm{d\hat{Q}}}{\rm{dt}} = \frac{i}{\hbar} \left[\hat{H}_{\rm{LC}}, \hat{Q}\right] = \omega_c Q_{\rm{zpf}} \left(c^{\dagger}+\hat{c}\right)\,,
\end{equation}
where $\hat{Q}=i Q_{\rm{zpf}}\left(\hat{c}^{\dagger} - \hat{c}\right)$ with $Q_{\rm{zpf}}= \sqrt{\nicefrac{\hbar C \omega_c}{2}}$ and $C$ is capacity. The magnetic part is quantized via
\begin{equation}
    \delta\hat{\textbf{m}}\left(\textbf{r},t\right) = \delta \textbf{m}\left(\textbf{r}\right) e^{-i\omega_m t}\hat{m} + h.c.\,,
\end{equation}
where we consider a splitting of the reduced magnetization, $\nicefrac{\textbf{M}}{\textbf{M}_S}$ with $\textbf{M}_S$ being the saturation magnetization, into a static magnetic ground-state plus the dynamic part on top, $\textbf{m}\left(\textbf{r},t\right) = \textbf{m}_0\left(\textbf{r}\right)+\delta \textbf{m}\left(\textbf{r},t\right)$. Inserting these expressions into Eq.~\eqref{eq:MWCoup} one obtains
\begin{align}
\label{eq:magnonmicrowavecoup}
    \hat{H}_{\rm{}Zeeman} &= \frac{\mu_0 \omega_c Q_{\rm{zpf}}}{4\pi} \left(\hat{c}^{\dagger}+\hat{c}\right) \nonumber \\
    &\left[\hat{m}^{\dagger}\int \rm{dr}^3 \delta \textbf{m}\left(\textbf{r}\right) \int_{U} \frac{I\left[dl \times \textbf{r}^{\prime} \right]}{|\textbf{r}^{\prime}|^3} + h.c.\right]\,,
\end{align}
which can be used to estimate the coupling strength of the resonant interaction between magnons and microwave photons described by the process $\hat{m}^{\dagger}\hat{c}+h.c.$
\subsection{Photoelastic Interaction}
Following the model described in Ref.~\cite{wolffBrillouinScatteringTheory2021}, we extend the analysis from a 2D waveguide picture to 3D resonators by integrating over all spatial dimensions instead of only over a resonator cross-section. There is a formal caveat in making this transition: the system modes $\psi_n=\begin{pmatrix}\bm{E}_n&\bm{H}_n\end{pmatrix}^T$ (where $\bm{E}_n$ and $\bm{H}_n$ are the electric and magnetic field distributions of the mode) are quasi-normal modes, such that the strict assumptions in Ref.~\cite{wolffBrillouinScatteringTheory2021} become approximate. In this framework, the general solution is a decomposition $\psi = \sum_n a_n \psi_n$, and the classical equations of motion for the complex amplitudes $a_n$ for the simplest magnetoelastic interactions take the form
\begin{equation}
    \left(\partial_t + \Gamma_n \right) a_n =\frac{i}{\mathcal{E}_n}\sum_{m} \left[e^{-i \Omega_\text{mech}t}\omega_m Q_{nm} b+ \text{c.c.}\right]a_m\,,
    \label{eq:optomech_mode_eq}
\end{equation}
where $b$ is the mechanical mode complex amplitude, $\omega_n$ and $\Omega_\text{mech}$ are the optical and mechanical mode frequencies, $\Gamma_n=\bra{n}\text{diag}(\sigma,0)\ket{n}$ is the matrix element corresponding to intrinsic Joule heating losses for a material with conductivity $\sigma$, $\mathcal{E}_n=\bra{n}\text{diag}(\epsilon' \varepsilon_0, \mu \mu_0)\ket{n}$ is the mode energy normalization, and the elements $Q_{nm}$ describe the strength of the coupling.

The nature of the interaction dictates that the lowest-order terms remaining in the expansion give the classical photon-pressure interaction part of the Hamiltonian presented in Eq.~\eqref{eq:Hamiltonian0}. Comparing Eq.~\eqref{eq:optomech_mode_eq} to the Heisenberg equation $d\hat a_n/dt = i/\hbar \left[\hat H, \hat a_n\right]$, one can immediately find the relation between the sought coupling constant and the matrix elements described above:
\begin{equation}
    g_{ab}^{(nm)} = \frac{\omega_m Q_{nm}}{\mathcal{E}_n}\,.
\end{equation}

In our geometry, the interaction elements $Q_{nm}$ are primarily determined by two mechanisms. The first mechanism is photoelasticity—changes in the material permittivity due to deformation, $\Delta\varepsilon_{ij}=-\varepsilon_r^2 p_{ijkl}\epsilon_{kl}$, where $p_{ijkl}$ is the photoelastic tensor and $\varepsilon_r$ is the real part of the permittivity. This interaction produces a matrix element
\begin{equation}
    \label{eq:photoelastic_cpl}
    Q^{(\text{PE})}_{nm} = \varepsilon_0 \varepsilon_r^2 \int_V d^3r \,  \sum_{ijkl}  E_i^{(n)\ast} E_j^{(m)} p_{ijkl} \epsilon_{kl}\,.
\end{equation}

The second mechanism is coupling due to boundary motion, which can be understood as arising from changes in optical mode frequencies due to deformation of the nanoresonator. This effect can be calculated perturbatively (see Ref.~\cite{johnsonPerturbationTheoryMaxwells2002}) using the following expression:
\begin{align}
    \label{eq:bound_cpl}
    Q^{(\text{MB})}_{nm} = \notag \\\int_\Gamma d^2r \, (\bm{u}^* \cdot \hat{\bm{n}})  &\left[(\varepsilon_a - \varepsilon_b) \varepsilon_0 (\hat{\bm{n}} \times \bm{e}^{(n)})^* \cdot (\hat{\bm{n}} \times \bm{e}^{(m)})  + \right.\notag\\
    &\left.(\varepsilon_b^{-1} - \varepsilon_a^{-1}) (\hat{\bm{n}} \cdot \bm{d}^{(n)})^* (\hat{\bm{n}} \cdot \bm{d}^{(m)}) \right]\,,
\end{align}
or alternatively evaluated directly by recomputing the optical modes in the deformed geometry.

\subsection{Numeric mode calculations}

In the course of calculating the interaction strengths we also need to know the mode distributions of the confined magnetic $\delta\bm{m}$, mechanical $\bm{u}$ and optical $\{\bm{E}, \bm{H}\}$ modes of the sample. We are using the axial symmetry of the system to simplify individual eigenproblems by considering an axial number $n_\varphi=0$, mainly limited by the coupling to microwave resonator --- the microwave wavelength is much larger than structure's size so that coupling to modes with $n_\varphi>0$ is greatly suppressed.

We use a custom proprietary finite-elements solver provided by partners at CEA for solving magnetic and elastic eigenproblems, for example in Ref.~\cite{bondarenkoMagnetoelasticConversionIntegrated2026}. The mechanical problem is solved using the linear isotropic elasticity equations \cite{landauTheoryElasticityVolume1986} using Young's modulus $E=\qty{198}{\giga\pascal}$, Poisson ratio $\nu=0.3$ and density $\rho=\qty{5170}{\kilo\gram\per\meter^3}$. Magnetic eigenmodes are similarly calculated for the Landau Lifshitz Gilbert equation \cite{brownMicromagnetics1963} linearized around the equilibrium vortex-state magnetization $\bm{m}_0$; the parameters for magnetic calculations are $A_{ex}=\qty{3.7}{\pico\joule\per\meter}$, $M_{s}=\qty{139}{\kilo\ampere\per\meter}$, and anisotropy is ignored.

For the optical quasi-normal (including radiation and material losses for our open resonator) mode calculation we implement the weak form described in Ref.~\cite{grudininFiniteelementModelingCoupled2012} modified by also including a perfectly match layer boundary to make the radiation losses calculation more accurate. Optically YIG has a transparency window in the near infrared extending all the way to the telecom c-band, although peak transparency is at higher wavelength \cite{dillonOpticalAbsorptionsRotations1959}. Within the band the refractive index is taken to be $n\approx 2.19$, at the lower end for typical optical microresonator materials. For estimating intrinsic losses we take $\epsilon''=4\times10^{-3}$. We neglect the optical dispersion (wavelength dependence) in the YIG for simplicity. The custom FENICs solver is set up for the $C_1$ continuous $\bm{H}$-field with mixed elements, Nédélec for $H_r,H_z$ components of field and Lagrange for the azimuthal component $H_\varphi$ to improve the solver's convergence.

\bibliographystyle{apsrev4-2}
\bibliography{TeleportationEntanglement.bib}
%
%
%
%
%
%
%

%
\end{document}